\documentclass[11pt]{article}
\pdfoutput=1
\usepackage{amsfonts,amsmath}
\usepackage{amssymb}
\usepackage{fancyhdr}
\usepackage{slashed}
\usepackage{graphicx}
\usepackage{subfigure}
\usepackage{color}

\usepackage{hyperref}
\usepackage[utf8]{inputenc}
\usepackage[titletoc]{appendix}

\def\hybrid{\topmargin -20pt    \oddsidemargin 0pt
        \headheight 0pt \headsep 0pt
        \textwidth 6.25in       
        \textheight 9.25in       
        \marginparwidth .875in
        \parskip 5pt plus 1pt   \jot = 1.5ex}

\hybrid

\def\baselinestretch{1.2}

\catcode`\@=11

\def\marginnote#1{}
\newcount\hour
\newcount\minute
\newtoks\amorpm
\hour=\time\divide\hour by60
\minute=\time{\multiply\hour by60 \global\advance\minute by-\hour}
\edef\standardtime{{\ifnum\hour<12 \global\amorpm={am}%
        \else\global\amorpm={pm}\advance\hour by-12 \fi
        \ifnum\hour=0 \hour=12 \fi
        \number\hour:\ifnum\minute<10 0\fi\number\minute\the\amorpm}}
\edef\militarytime{\number\hour:\ifnum\minute<10 0\fi\number\minute}

\def\draftlabel#1{{\@bsphack\if@filesw {\let\thepage\relax
   \xdef\@gtempa{\write\@auxout{\string
      \newlabel{#1}{{\@currentlabel}{\thepage}}}}}\@gtempa
   \if@nobreak \ifvmode\nobreak\fi\fi\fi\@esphack}
        \gdef\@eqnlabel{#1}}
\def\@eqnlabel{}
\def\@vacuum{}
\def\draftmarginnote#1{\marginpar{\raggedright\scriptsize\tt#1}}

\def\draft{\oddsidemargin -.5truein
        \def\@oddfoot{\sl preliminary draft \hfil
        \rm\thepage\hfil\sl\today\quad\militarytime}
        \let\@evenfoot\@oddfoot \overfullrule 3pt
        \let\label=\draftlabel
        \let\marginnote=\draftmarginnote
   \def\@eqnnum{(\theequation)\rlap{\kern\marginparsep\tt\@eqnlabel}%
\global\let\@eqnlabel\@vacuum}  }

\def\preprint{\twocolumn\sloppy\flushbottom\parindent 2em
        \leftmargini 2em\leftmarginv .5em\leftmarginvi .5em
        \oddsidemargin -.5in    \evensidemargin -.5in
        \columnsep .4in \footheight 0pt
        \textwidth 10.in        \topmargin  -.4in
        \headheight 12pt \topskip .4in
        \textheight 6.9in \footskip 0pt
        \def\@oddhead{\thepage\hfil\addtocounter{page}{1}\thepage}
        \let\@evenhead\@oddhead \def\@oddfoot{} \def\@evenfoot{} }

\def\numberbysection{\@addtoreset{equation}{section}
        \def\theequation{\thesection.\arabic{equation}}}

\def\underline#1{\relax\ifmmode\@@underline#1\else
        $\@@underline{\hbox{#1}}$\relax\fi}

\def\titlepage{\@restonecolfalse\if@twocolumn\@restonecoltrue\onecolumn
     \else \newpage \fi \thispagestyle{empty}\c@page\z@
        \def\thefootnote{\fnsymbol{footnote}} }

\def\endtitlepage{\if@restonecol\twocolumn \else \newpage \fi
        \def\thefootnote{\arabic{footnote}}
        \setcounter{footnote}{0}}  

\catcode`@=12
\relax

\def\figcap{\section*{Figure Captions\markboth
        {FIGURECAPTIONS}{FIGURECAPTIONS}}\list
        {Figure \arabic{enumi}:\hfill}{\settowidth\labelwidth{Figure
999:}
        \leftmargin\labelwidth
        \advance\leftmargin\labelsep\usecounter{enumi}}}
 \relax
\def\tablecap{\section*{Table Captions\markboth
        {TABLECAPTIONS}{TABLECAPTIONS}}\list
        {Table \arabic{enumi}:\hfill}{\settowidth\labelwidth{Table
999:}
        \leftmargin\labelwidth
        \advance\leftmargin\labelsep\usecounter{enumi}}}
 \relax
\def\reflist{\section*{References\markboth
        {REFLIST}{REFLIST}}\list
        {[\arabic{enumi}]\hfill}{\settowidth\labelwidth{[999]}
        \leftmargin\labelwidth
        \advance\leftmargin\labelsep\usecounter{enumi}}}
 \relax
\makeatletter
\newcounter{pubctr}
\def\publist{\@ifnextchar[{\@publist}{\@@publist}}
\def\@publist[#1]{\list
        {[\arabic{pubctr}]\hfill}{\settowidth\labelwidth{[999]}
        \leftmargin\labelwidth
        \advance\leftmargin\labelsep
        \@nmbrlisttrue\def\@listctr{pubctr}
        \setcounter{pubctr}{#1}\addtocounter{pubctr}{-1}}}
\def\@@publist{\list
        {[\arabic{pubctr}]\hfill}{\settowidth\labelwidth{[999]}
        \leftmargin\labelwidth
        \advance\leftmargin\labelsep
        \@nmbrlisttrue\def\@listctr{pubctr}}}
 \relax
\makeatother
\newskip\humongous \humongous=0pt plus 1000pt minus 1000pt

\newif\ifdtup

\relax

\def\be{\begin{equation}}
\def\ee{\end{equation}}
\def\ba{\begin{eqnarray}}
\def\ea{\end{eqnarray}}

\def\r{\rho}
\def\a{\alpha}

\def\b{\beta}

\def\g{\gamma}

\def\d{\delta}

\def\e{\epsilon}

\def\th{\theta}

\def\m{\mu}

\def\Om{\Omega}
\def\l{\lambda}

\def\s{\sigma}

\def\vphi{\varphi}
\def\cN{{\cal N}}

\def\rt{\tilde r}
\def\cM{{\cal M}}

\def\cL{{\cal L}}

\def\cD{{\cal{D} } }

 \def\cB{{\cal B}} 
\def\cD{{\cal D}}  \def\cF{{\cal F}}
 \def\cH{{\cal H}} 
  \def\cL{{\cal L}}
\def\cM{{\cal M}} \def\cN{{\cal N}} \def\cO{{\cal O}}
  
 \def\cT{{\cal T}} 
  
 \def\cZ{{\cal Z}}

\newcommand{\prt}[1]{{\left( {#1} \right)}}

\def\no{\noindent}

\def\IR{\relax{\rm I\kern-.18em R}}

\def\pp{\partial}

\newcommand{\ff}{\frac}

\def\IR{\relax{\rm I\kern-.18em R}}
\def\IL{\relax{\rm I\kern-.18em L}}

\def\inv{^{\raise.15ex\hbox{${\scriptscriptstyle -}$}\kern-.05em 1}}

\def\cM{{\cal M}}

\def\cL{{\cal L}}

\def\bea{\begin{eqnarray}}
\def\eea{\end{eqnarray}}
\newcommand{\eq}[1]{(\ref{#1})}
\def\nn{\nonumber}

\newcommand{\la}[1]{\label{#1}}

\def\a{\alpha}      
\def\b{\beta}       
\def\g{\gamma}    
\def\d{\delta}    
\def\e{\epsilon}

\def\l{\lambda} 
\def\m{\mu}

\def\r{\rho}
\def\s{\sigma}  
\def\t{\tau}
\def\th{\theta}

\definecolor{markcolor2}{rgb}{1,0,0}

\definecolor{markcolor3}{rgb}{0,1,0}

\begin{document}

\renewcommand{\theequation}{\thesection.\arabic{equation}}
\csname @addtoreset\endcsname{equation}{section}

\newcommand{\beq}{\begin{equation}}
\newcommand{\eeq}[1]{\label{#1}\end{equation}}
\newcommand{\ber}{\begin{eqnarray}}
\newcommand{\eer}[1]{\label{#1}\end{eqnarray}}
\newcommand{\eqn}[1]{(\ref{#1})}
\begin{titlepage}

\begin{center}

~
\vskip 1 cm

{\Large
\bf Baryons under Strong Magnetic Fields \\or\\ in Theories with Space-dependent $\theta$-term
}

\vskip 0.7in

 {\bf Dimitrios Giataganas}
 \vskip 0.1in
 {\em
     Physics Division, National Center for Theoretical
   Sciences, \\
  National Tsing-Hua University, Hsinchu, 30013, Taiwan
 \\\vskip .1in
 {\tt ~  dimitrios.giataganas@cts.nthu.edu.tw
 }\\
 }

\vskip .2in
\end{center}

\vskip .8in

\centerline{\bf Abstract}

Baryonic states are sufficiently complex to reveal physics that is hidden in the mesonic bound states.
Using gauge/gravity correspondence we study analytically and numerically baryons in theories with space-dependent $\theta$-term, or theories under strong magnetic fields. Such holographic studies on baryons are accommodated in a generic analytic framework we develop for anisotropic theories, where their qualitative features are common  irrespective of the source that triggers the anisotropy. We find that the distribution of the quarks forming the state, depends on the angle between the baryon and the anisotropic direction. Its shape is increasingly elliptic with respect to the strength of the field sourcing the anisotropy, counterbalancing the broken rotational invariance on the gluonic degrees of freedom. Strikingly, the baryons dissociate in stages  with a process that depends on the proximity of the quarks to the anisotropic direction, where certain quark pairs abandon the bound state first, followed by the closest pairs to them as the temperature increases. This observation may also serve as a way to identify the nature of certain  exotic states. Finally, we investigate holographic baryons with decreased number of quarks and explain why in theories under consideration the presence of anisotropy does not modify the universal stability condition in contrast to the usual trend.

\no
\end{titlepage}
\vfill
\eject


\noindent


\def\baselinestretch{1.2}
\baselineskip 19 pt
\noindent


\setcounter{equation}{0}

\section{Introduction, Methodology and Results}

Quarkonium physics is fundamental to reveal properties of the QCD and its dynamics.  The early quarkonium studies were based on the non-relativistic potentials \cite{nrpoten1,nrpot2}, while the present formulation is in the context of the effective field theories of QCD,  lattice-QCD \cite{Brambilla:2004wf,Brambilla:2010cs,Bali:2000gf,Gray:2005ur} and of the gauge/gravity duality \cite{adscft1,CasalderreySolana:2011us} where the focus is mostly on qualitative but still insightful physics. The research of meson resonances, hybrids and multiquark bound states has been a very active field for many years, while recently there is an increasing interest on the quark bound states and baryons due to new experimental and theoretical developments.  Baryons, may be relatively simple, but are sufficiently complex to reveal physics  hidden in the mesonic bound states, making them worthy to study.

From theoretical point of view, the computations on baryonic observables are much more demanding than the mesonic ones. For example, in lattice-QCD the baryonic correlation functions have increased statistical noise,  while the computation of the quark-disconnected diagrams suffer from large statistical fluctuations. Although the progress in lattice-based techniques is impressive, there exist certain obstacles and several known baryonic structural properties have not been reproduced yet. Light baryon resonances is another puzzling area for the strong QCD dynamics. Quark models and lattice-QCD calculations predict excited states which have not been observed, questioning the validity of the methods predicting the existence of such states or perhaps the feasibility to identify them in the experimental data \cite{Capstick:2000qj,Edwards:2011jj,Crede:2013sze}. There are similar challenges related to the existence of certain resonances in heavy and light baryon sector and the baryon chiral dynamics. While their complete listing is beyond the scope of the current work, it is worthy to mention the recent exciting discoveries and developments on the exotic hadrons \cite{Ali:2017jda,Olsen:2017bmm,Chen:2016qju,Esposito:2016noz,Lebed:2016hpi,Guo:2017jvc,Chen:2016spr}.  The possibility of the exotic hadrons was anticipated in the very early days of the introduction of quarks \cite{gellmannquarks,Zweig:1981pd} and the discovery of the exotic hadron $X(3872)$ in 2003 \cite{Aubert:2004ns,Choi:2007wga} came as a surprise and opened new directions of research on the  QCD bound states. Today there are more than thirty observed candidates of exotic hadrons, where the possible structures of such bound states have been proposed to be a type of hadronic molecules, hybrid mesons, diquarks, tetraquarks and pentaquark baryons.  The fundamental question is whether or not exist a unique physical picture sufficient to explain all the new experimental data on production mechanisms, masses, decay modes and rates, capable of identifying the exact type of the bound state observed from the options listed above. So far such a unique collective framework has not been found and it is under debate if it exists. 

Motivated by all these developments and the continuous research interest on the topic, we study holographic baryons in theories under magnetic fields or
in theories with space dependent $\theta$-terms. The study of baryons under such anisotropic theories reveals new interesting properties,  while our  findings indicate that most of the qualitative effects on baryons are sensitive to the presence of the anisotropies than the source generating them.

Strongly coupled systems in presence of strong external fields are not rare, for example strong short-lived fields occur in the heavy ion collisions of almost constant strength during the plasma's life, which couple directly to the charged fermions  and indirectly to gluons.  In the presence of magnetic field in QCD,  several peculiar phenomena  appear when anisotropies are present \cite{Kharzeev:2012ph,Miransky:2015ava}. What makes such studies even more interesting from the point of view of holography, is that most observables and phenomena   depend mostly on the fact that the anisotropies are present, irrespective of the source triggering them. This has been demonstrated initially by studying heavy quark observables \cite{Giataganas:2012zy} and showing how their properties depend on the anisotropies of the theory. Moreover, the phase diagram of QCD matter is expected to be affected by phenomena that exist in static equilibrium state in presence of magnetic fields,  the so called inverse magnetic catalysis where both the chiral symmetry restoration and deconfinement occur at lower temperatures in the presence of an external magnetic field \cite{Bali:2011qj,Bali:2011uf,DElia:2012ems,Preis:2012fh}. Recent holographic studies brought a new twist in the story indicating that a similar phenomenon takes place even in plasmas with uncharged particles where the isotropy is broken, suggesting the possibility that the anisotropy by itself is the main cause of inverse magnetic catalysis \cite{Giataganas:2017koz}. Other interesting holographic studies in  presence of the magnetic fields include the lattice ground states \cite{Bu:2012mq},  where various instabilities occur, as in \cite{BallonBayona:2012wx}, while relevant phenomena in the presence of fundamental matter are reviewed in \cite{Bergman:2012na}. Our main focus in this work is on the study of the effect of the anisotropies generated by such strong fields on the multiquark bound states and more particularly on baryons.

An alternative source of anisotropies in the strongly coupled theories is the presence of space-dependent $\theta$-term. Constant coupling $\theta$-terms: $\theta~ \mbox{Tr} F\wedge F$, are induced in QCD by instanton effects, where the value of the topological parameter is predicted by experimental-theoretical analysis to be very low \cite{Afach:2015sja,Baluni:1978rf}. From the theoretical perspective such studies require non-perturbative techniques, where lattice can be useful but with limitations due to the sign problem \cite{Vicari:2008jw}, while there is also progress by using the gauge/gravity techniques e.g. \cite{Arean:2016hcs}. In a different framework, the $\theta$-terms in the field theories have been introduced recently to provide a consistent top-down  scheme to generate anisotropic gauge/gravity dualities. The idea of space-dependent or time-dependent couplings has been used in the past in several contexts, for example the former by generalizing half-BPS Janus configurations in $\cN=4$ sYM theory to show their relation to the existence of three-dimensional Chern-Simons theories \cite{Gaiotto:2008sd} or more recent generalizations \cite{Choi:2017kxf}, and the latter to construct time-dependent IIB supergravity solutions, for example \cite{Chu:2006pa}. Relevant to the purposes of our paper, more recently D3-D7 back-reacted supergravity solutions have been constructed with initial motivation to study  the pure Chern-Simons gauge theory. The solutions are non-supersymmetric, may be exact Lifshitz-like or may interpolate between the $AdS$ in the UV boundary and the Lifshitz-like in the IR
boundary \cite{Azeyanagi:2009pr,Mateos:2011tv,Giataganas:2017koz}. The flow is triggered by a non-zero $\theta$-parameter, that depends linearly on a spatial coordinate $x_3$. The connection of the $\theta$-term to the D7 branes comes when we relate the $\theta$ coupling to the axion term $\chi$ in the gravity dual theory, through the complexified coupling constant $\t$. The Hodge dual of the axion field strength in 10 dimensions gives $d C_8\sim \star d\chi$ with the non-zero component $C_{t x_1 x_2 M^5}$ to live on the rest of the spatial dimension and the internal space $M^5$. These are the directions that the backreacting D7-branes extend. Such theories turned out to be very interesting, and have been studied extensively initially with heavy quark probes in \cite{Giataganas:2012zy} and \cite{Chernicoff:2012iq}, while it has been found that they violate the well known shear viscosity over entropy density ratio for certain tensor components \cite{Rebhan:2011vd,Jain:2014vka,Giataganas:2017koz}.  The derivation of such confining theories  and their phase transitions has been done recently \cite{Giataganas:2017koz} opening the window for further   studies in the confined region.

\subsection{Methodology and Results} \label{section:method}

In our work we consider holographic baryons in anisotropic theories,  where it turns out that our qualitative findings are insensitive on the source triggering the anisotropy.
The construction involves the Witten baryon vertex on the internal space, acting as a heavy bag to absorb the $N_c$ units of flux, generated by the $N_c$ fundamental strings initiating from the boundary representing $N_c$ number of heavy quarks, ending on the brane vertex \cite{Witten:1998xy,Gross:1998gk}. The vertex is represented by a D5-brane, sitting at an appropriate point in the AdS space, and its dynamics is realized by the Dirac-Born-Infeld action plus the appropriate Wess-Zumino term. The $k$ number of strings are hanging down from the boundary to meet the D5-brane. Following symmetry arguments and previous relevant configurations  the boundary quarks tend to be placed in the most symmetric way, since in isotropic theories this ensures the stabilization of the state.  When the number of strings is $k=N_c$ the baryon is stabilized by the mass of the brane counterbalancing the tension of the strings, while on the spacetime plane we need to equate the tension forces of the strings. When $k<N_c$ the conservation of charges and the stability of the baryon is ensured with the presence and the tension of the $N_c-k$ strings between the brane and the deep IR of the space \cite{Brandhuber:1998xy}. A simplified explanation for the brane is that it acts as a baryon junction. However the vertex's position is completely determined by the size of the baryon on the boundary and the external fields, therefore a more accurate picture for the vertex is that it represents the resummed effect of the quarks on the gluonic fields of the state. Our fundamental strings are located at the same point in the $S^5$ and  for the purposes of our work which initiates the baryon studies on the anisotropic spaces, we ignore the interaction between them. Their dynamics are described by the Nambu-Goto action, while the inclusion of their interactions is not expected to change qualitatively our results. Here we follow the construction of \cite{Witten:1998xy,Gross:1998gk,Brandhuber:1998xy}, nevertheless we mention that a way to capture such string interactions is via the Born-Infeld action, as in the cases of \cite{Imamura:1998gk,Callan:1998iq,Craps:1999nc,Gomis:1999xs}  where baryonic branes are BPS  and several technical simplifications occur in contrast to our case. Moreover, we point out that in wide class of theories satisfying a condition related to the existence of preserved quantities under certain T-dualities, the Dp-brane $k$-string observables are mapped to the fundamental string observables \cite{Giataganas:2015ksa}. In such theories the interaction of the strings is not expected to affect our results significantly since only constant rescalings occur between the energy of the brane and the energy of a string.

The study of baryons has been done so far in several theories and with different tools. These include: holographic baryons described by soliton  configurations \cite{Hashimoto:2008zw} and various approximations of it \cite{Bolognesi:2013nja}, baryons from oblate instantons \cite{Rozali:2013fna}, baryons moving in the plasma wind \cite{Athanasiou:2008pz}, heavy and exotic states from instanton in holographic QCD \cite{Liu:2017xzo}, several types of multiquark bound states \cite{Sonnenschein:2016pim,Sonnenschein:2016ibx}, baryons in presence of medium \cite{Seo:2008qc}, stability of baryons \cite{Sfetsos:2008yr},  studies with certain bottom-up potentials \cite{Andreev:2008tv}, studies making use of large $N_c$ approximations  \cite{Kim:2017khv}.

\subsubsection{The Theory-Independent Approach}

In our current work we focus on the effect of the anisotropies triggered by several sources, like the magnetic fields or the presence of space-dependent $\theta$-terms.  The qualitative effects on the baryonic states are independent of the nature of the anisotropic source, and therefore are expected to hold for a wide range of anisotropic theories. We start by considering a generic class of gravity dual backgrounds with certain field strengths and perform the analysis in a theory-independent way. We derive the necessary stability condition for the baryon configuration, along the holographic and the spatial directions  by equating the total string tension with the mass of the brane vertex in the radial direction and the string tensions along the spatial directions. In practice, these conditions are derived by the requirement of cancelling the boundary terms of the string and brane actions, which establish a relation of the position of the brane in holographic direction with respect to the size of the baryon. Then the energy of the baryon is derived with respect to its size  and the shape of the quarks at the boundary. We observe that the properties of the baryon depend on the angle between the plane where the quarks are located and the direction of the anisotropy, e.g. the magnetic field vector. There are two extremal ways to place the quarks of the baryon in such cases, enough to extract all the interesting information: on the anisotropic plane and on the transverse plane with the rotational invariance. When the baryon lives on the rotational invariant plane the quarks are distributed homogeneously on a cycle. In the anisotropic plane the distribution of the quarks is deformed to elliptic-like shape, in a way to counterbalance the spatial anisotropy on the geometry, i.e. in theories with space-dependent theta-term and prolate dual geometries, we get oblate quark distributions. This has significant implications to the way that the baryons dissociate: increase of the anisotropy leads to dissociation of the baryon's quarks in stages, depending on the angle of the quarks with respect to the anisotropic direction, where certain pairs abandon the bound state first, followed by the closest pairs to them as the temperature increases. Therefore, providing a unique feature of the effects of anisotropies on the baryonic states, having potential applications to multi-quark bound states, for example the exotic baryons. Exotic states that are strongly coupled are expected according to our study to dissociate in stages with a particular pattern, and as a result our findings serve potentially as a test to identify the nature of the exotic states. We point out, that this is a model-independent observation and independent of the way that the anisotropy is triggered, as long as the natural assumptions considered hold.

Still in the generic framework we then study certain aspects of baryons with $k<N_c$ quarks. We focus on the effect of the anisotropy on the universal relation of the stability condition involving the number of the boundary quarks with respect to the number of colors $5 N_c/8<k\le N_c$, which has been found to hold in several isotropic strongly coupled theories \cite{Brandhuber:1998xy,Giataganas:2012vw}. The examination of the universality relations on anisotropic theories is an interesting subject on its own, due to the well know ability of the presence of anisotropies to violate   universal results of isotropic theories. One of the most known relations is the parametric violation of particular components of the shear viscosity over entropy density ratio \cite{Rebhan:2011vd,Jain:2014vka,Giataganas:2017koz}. Similar violation occurs to the universal isotropic inequality  of the transverse and longitudinal Langevin coefficients of the heavy quark diffusion \cite{Giataganas:2013hwa,Gursoy:2010aa}, with respect to its motion in the strongly coupled anisotropic plasma \cite{Giataganas:2013hwa,Giataganas:2013zaa} \footnote{Notice that the general formalism for the heavy quark diffusion developed in \cite{Giataganas:2013hwa,Giataganas:2013zaa}, apply in straightforward way to certain limits on theories with magnetic fields and has been used in such environments already, for example in \cite{Dudal:2018rki}.}. Moreover, the screening length for the quarkonium mesons moving through an anisotropic strongly coupled plasma, scales in a different way with respect to the Lorentz factor $\g$ compared to the isotropic result \cite{Chernicoff:2012bu} \footnote{A review of strongly coupled anisotropic observables can be found in \cite{Giataganas:2013lga}.}. Therefore, the aforementioned observations link the presence of anisotropy with the weakening of applicability of the universal relations. In contrast, here we find that the   stability condition is insensitive to the presence of anisotropy in theories with AdS asymptotics.

\subsubsection{The Application on Theories with Space-Dependent $\theta$-terms or Magnetic Fields}

Next we apply our formalism to particular theories and perform a semi-analytical and extensive numerical analysis. We find numerically the gravity dual RG flows in IIB supergravity with space-dependent axion term in finite temperature for the whole range of anisotropies, and analytically the gravity dual background in the low anisotropy regime. In the latter we show  how the quark distribution is deformed in terms of integrands, under the effect of the space-dependent $\theta$-term. By performing the full numerical analysis we find the relation of the size and shape of the baryon with respect to a) the strength of anisotropy in the theory, b) the angle between the plane that the quarks are distributed and the anisotropic direction, and c) the position of the vertex brane in the bulk. Quarks distributed on the transverse to the anisotropic direction plane have cyclic distribution, while the increase of the anisotropy leads to the reduction of the size of the baryons, in order to retain stability. When the quarks are distributed on the anisotropic plane, the pairs along the anisotropy getting closer compared to the ones in the transverse direction, resulting to an elliptic-shaped distribution. By computing the
energy of the baryons we find that they dissociate in stages, where the quarks along the anisotropic directions abandon the bound state first followed by the closest pairs to them as the anisotropy increases.  In the theories under consideration, after certain high values of anisotropy the baryons show weak dependence on it. We elaborate in detail on this observation and we identify the reasons that this happens.

Then we show how the aforementioned findings, carry on qualitatively for theories with magnetic fields. This is done by studying the brane and string dynamics in presence of these fields, and showing that the main contributions of the effects on the magnetic field on the holographic baryonic states are induced by the anisotropic geometry. Constant magnetic fields  contribute only through the modification of the geometry to the string Nambu-Goto equations for our system, while on the brane may induce constant shifts. This justifies that the baryons under strong magnetic fields dissociate in stages as we have described above, and the rest of the qualitative  properties we have found still hold for this case.

We point out that our observations have potential applications to exotic baryons. Although, we do not study the holographic exotics explicitly, the qualitative features of our holographic studies are expected to carry on such multiquark bound states, since the first principles of the holographic construction and computations are the same. This implies that under strong fields, strongly coupled exotic states bounded in the same way with the conventional baryons are expected to dissociate in stages with a particular pattern. This can serve as a potential test to identify the nature of the exotic states, i.e. molecule states will not have this property.

The plan of the paper is the following. In section \ref{section:generic} we study the holographic baryon in a wide class of anisotropic theories  with external fields by imposing natural assumptions. Of particular interest is how the distribution of the baryonic quarks is affected by the strength and direction of the anisotropy presented in subsection \ref{section:shape}. In section \ref{section:theta} we study in depth, analytically and numerically the baryon in theories with space-dependent $\theta$-term, where several subsections included. In section \ref{section:mag}, we show that in the presence of the magnetic fields,
the main contributions to the baryonic properties come from the space-time anisotropic geometry. Then in section \ref{section:univ} we study the stability condition for baryons with $k< N_c$ quarks in anisotropic spaces. We briefly present the final remarks in section \ref{section:final}, and support our main text with an appendix.

\section{Baryon in Theories with Strong Fields: The Formalism} \label{section:generic}

The analysis of this section and the section  \ref{section:mag} is carried out for general theories. The gravity dual theory we need to consider has reduced rotational symmetry in the spatial volume. The baryon constructed in those theories have $k=N_c$ fundamental strings initiating from the boundary of the space where the heavy quarks are located, and dive to the interior of the space ending on a D5-brane wrapped on the internal space $\cM^5$. The rotational symmetry of the background is reduced, so we have to determine the energetically preferable positions of the $N_c$ external quarks at the boundary by studying the dynamics of the system. In the special case of the isotropic theories the distribution can be determined to be cyclic, consistent with symmetry arguments of the state.

The dual field theory considered may be at a nonzero temperature, so its gravity background is allowed to contain  a black hole with an anisotropic horizon. We do not consider NS-NS field in this section although such complications are not expected to add qualitatively  new features to the analysis of the baryon configurations and we study this case at a later section \ref{section:mag}.

The metric of such  theory in string frame reads
\be\la{metric1}
ds^2=g_{00}(r) dt^2+g_{11}(r)\prt{dx_1^2+dx_2^2}+g_{33}(r) dx_3^2 +g_{rr}(r)dr^2+g_{\th\th}(r) d\cM_5^2~,
\ee
where a non-trivial dilaton $\phi(r)$ is considered among other scalar fields like the axion, which do not couple directly to string dynamics. The $\cM_5$ is a five-dimensional internal compact manifold. The properties of the metric are
\be
g_{00}(r_h)=0~,\qquad \lim_{r\to \infty} g_{11}(r)=\lim_{r\to \infty} g_{33}(r)= \infty~,
\ee
where $r_h$ is the horizon of the anisotropic black hole and $r$ is the holographic direction of the space.

The baryon configuration involves two different extended objects. The $N_c$ fundamental strings from the boundary ending on the baryon vertex, and the vertex Dp-brane wrapping the internal compact space. The positions of the quarks on the boundary denoted be $\vec{x}(\infty)^k$ with $k=1,\ldots,N_c$ and they are localized in the compact manifold $\cM_5$. The D5 brane vertex is located at $(\vec{x}_v,r_v)$, where we shift the coordinate system  such that $\vec{x}_v=0$. Having a system of fundamental strings and a brane, the total action describing it  consists of the Nambu-Goto (NG) and the Dirac-Born-Infeld (DBI) actions
\be
 S_{\rm{tot}}=S_{D5}+\sum_{k=1}^{N_c} S_{F1}^k~.
\ee
We may consider the radial gauge and without loss of generality to parametrize the system as $t=\t,~r=\s,~x_1=x_1(\s),~x_2=0,~x_3=x_3(\s)$, for each of the $k$ strings. The analysis below is for  localized  string along the $x_2$ plane, but without any complication we can generalize to contain an extension along $x_2$ to apply the  formulas in a straightforward way. One may also ask why we choose to localize the string on the internal space $\cM$. Non-localized strings along the internal space cost additional energy and for the purposes of our current work this is not the physically preferable configuration.

Having clarified the choice of the parametrization, the total action takes the form
\bea\nn
 &&S_{\rm{tot}}=\frac{\cT}{2 \pi \a'}\sum_{k=1}^{Nc}\int_{r_v}^{\infty}dr\sqrt{-g_{00}\prt{g_{rr}+g_{ii} x_{i}{}_{(k)}'(s)^2}}+\frac{\cT \rm{vol}_5}{\prt{2\pi}^5\a'^3 g_s} \sqrt{-g_{00}(r_v) g_{\th\th}(r_v)^5}e^{-\phi(r_v)}\\\la{action1}
 &&:=\frac{\cT}{2 \pi \a'}\sum_{k=1}^{Nc} \int_{r_v}^{\infty}dr\sqrt{\cD_{1}{}_{(k)}}+\frac{\cT \rm{vol}_5}{\prt{2\pi}^5\a'^3 g_s} \cD_2~,
\eea
where $i=1,3$, the  $\rm{vol}_5$ is the volume of the compact manifold  and the second term is evaluated at $r=r_v$ the position of the D5 brane.
The configuration has two constants of motion for each string $k$ of the $N_c$ strings
\be\la{qi}
Q_{i}{}_{(k)}= -\frac{g_{00} g_{ij} x_j'{}_{(k)}(s)}{\sqrt{\cD_{1}{}_{(k)}}}~.
\ee
The system can be solved analytically for $x_i'$, for each of the $k$ strings giving
\bea \la{xisol1}
&&x_1'^2(s)=-\frac{Q_1^2 g_{33} g_{rr}}{g_{11}\prt{Q_3^2 g_{11}+Q_1^2 g_{33}+g_{00} g_{11} g_{33}}}~,\\
&&x_3'^2(s)=-\frac{Q_3^2 g_{11} g_{rr}}{g_{33}\prt{Q_3^2 g_{11}+Q_1^2 g_{33}+g_{00} g_{11} g_{33}}}~,\la{xisol3}
\eea
where for presentation reasons we have dropped the notation labeling the string $k$.
Notice the symmetry in the last two equalities making them equivalent in the case of isotropic space $g_{11}=g_{33}$. In all other cases, our equations indicate that the boundary distribution is not cyclic as long as a homogeneous quark distribution is considered.

\subsection{Stability of the State}

To ensure the stability of the baryon we impose further conditions by equating the forces along the spatial plane and the radial direction as in \cite{Athanasiou:2008pz}. The variation of the total action along the spatial plane, gives the "E-L" terms proportional to the Euler-Lagrange equations plus  boundary terms at the location of the brane
\be\la{vara1}
\int \d\cL_{tot} d r=-\sum_{k=1}^{Nc}\prt{\frac{\pp\cL_{F_1}}{\pp x'^i_{(k)}}\d x^i_{(k)} +\frac{\pp\cL_{F_1}}{\pp r'_{(k)}}\d r_{(k)}}+ \frac{\pp\cL_{D5}}{\pp r}\d r+ \mbox{E-L}~.
\ee
Therefore the forces along the spatial plane are expressed with the constants of motion \eq{qi} and their sum needs to be zero as
\be\la{forcex1x3}
\sum_{k=1}^{N_c} \prt{Q_1{}_{(k)}+Q_3{}_{(k)}}\big|_{r=r_v}=0~.
\ee
The above conditions ensures that the configuration has no force on $x_1$ and $x_3$ plane. It is equivalent of saying that each quark and its antipodal partner hit the brane vertex with the same angle. In the continuous limit we may generalize this argument and ensure that the above condition is satisfied by requiring that all the strings hit the brane with an angle $x'(\s)$. To describe the quark distribution let us introduce an azimuthal angle $\varphi$ in the $(x_1,x_3)$ plane then $x'(\s)$ is defined by $x_1'(\s):=x'(\s) \sin\vphi$, $x_3'(\s):=x'(\s) \cos \vphi$ at the position of the brane.

To ensure stability in the holographic direction we find from \eq{vara1}
\be\la{forcedp1}
\sum_{k=1}^{N_c} \frac{-g_{00} g_{rr}}{\sqrt{\cD_{1}{}_{(k)}}}\bigg|_{r=r_v}= \frac{ N_c  }{4} \frac{\pp\cD_2}{\pp r}\bigg|_{r=r_v}~.
\ee
This is understood as the condition equating the tension force of strings on the brane pointing to the boundary, with the gravitational force on the mass of the D5 brane. The RHS comes with a factor  $\rm{vol}_5 R^4/\prt{\prt{2\pi}^5\a'^3 g_s}$, where the radius of the space $R$, is already extracted from the expression $D_2$ to simplify the constants.
The equation \eq{forcedp1} is nothing but the radial force of the fundamental strings which is equal in magnitude to their Hamiltonian $\cH$, while of the right hand side is the radial derivative of the D5 brane action.
Alternatively it may be  written  as
\be
-\sum_{k=1}^{N_c}  \cH_{F_1}{}_{(k)}= \frac{\partial S_{D5}}{\partial r} .
\ee
At the continuous quark limit where the sum is normalized as $N_c\int_0^{2\pi}d\phi/(2\pi)$, the condition \eq{forcedp1}  takes the following form
\be\la{forcedp2}
\int_0^{2\pi}\frac{d\varphi}{2\pi}\frac{g_{00} g_{uu}}{ \sqrt{-g_{00}\prt{g_{rr}+ x'(s)^2  \prt{g_{11} \sin^2\vphi+g_{33}  \cos^2\vphi } }}}\bigg|_{ r_v}= \frac{e^{-\phi}g_{\th\th}^{3/2}\prt{5 g_{00} g_{\th\th}'+g_{\th\th}\prt{g_{00}'-2g_{00}\phi'}}}{8 \sqrt{-g_{00}}}\bigg|_{ r_v}~.
\ee
To summarize, our generic conditions \eq{forcex1x3} and \eq{forcedp1} ensure a stable configuration and no-force condition in the spatial and holographic directions. These conditions should be imposed on the brane-string configuration satisfying the equations \eq{xisol1} and \eq{xisol3} which ensure a  minimized total action.

\subsection{Energy of the state}  \label{section:energy}

The contributions to the energy of the baryon come from the $N_c$ fundamental strings and the Dp-brane subtracting the infinite masses of the quarks. The latter is just the length of the infinite string
\be
 E_m=\frac{1}{2\pi \a'} \int_{r_h}^{\infty} d\s \sqrt{ -g_{00} g_{rr}}~.
\ee
The rest comes from the minimization of the NG and the DBI actions \eq{action1}. Therefore the total regularized energy can be found as
\bea\nn
E_{\mbox{baryon}}&=&\frac{N_c}{2 \pi \a'} \Bigg[ \int_{r_v}^{\infty}d\s \prt{\frac{1}{N_c}\sum_{k=1}^{Nc}\sqrt{-g_{00}\prt{g_{rr}+g_{ij} x_{j}{}_{(k)}'(s)^2}} -\sqrt{ -g_{00} g_{rr}}}\\\la{enea1}
&&+ \int_{r_v}^{r_h}d\s\sqrt{-g_{00}g_{rr}}+\frac{1}{4}e^{-\phi(r_v)}\sqrt{-g_{00}(r_v)g_{\th\th}(r_v)^5}\Bigg]~,
\eea
where the radius $R$ has been taken out from the metric element expressions.
Notice that the brane contributes only a constant positive term proportional to the number of colors $N_c$. This is universal behavior irrespective of the space and the theory we are examining as long as we consider wrapping branes. More involved brane configurations to play the baryon's vertex role are in principle allowed, however these are expected to have higher energies.

\subsection{Quark Distribution in Presence of Strong Fields}  \label{section:shape}

There are two extremal ways the heavy baryon can be placed in the theory, while the rest cases consist of a decomposition of these two. The most interesting is when the quarks lie along the $(x_1,x_3)$ anisotropic plane where the backreaction of the field breaks the rotational invariance. The gluons of the theory have been affected by the external fields resulting to a quark distribution either non-homogeneous or non-symmetric/non-cyclic, as hinted already by the equations \eq{xisol1}. Alternatively the baryon can be placed with the quarks being distributed along a plane with rotational symmetry. In this case the distribution of the quarks is symmetric around a point, while the energy and the properties of the bound state will be modified due to the presence of the external field.



\subsubsection{Bending Strings and Quark Distribution for a Baryon on the Anisotropic Plane}

Let us  analyze the non-trivial baryon where the $N_c$ quarks are distributed in the plane $(x_1,x_3)$ with the reduced rotational invariance. The  distribution on the boundary is found by  integrating \eq{xisol1} and \eq{xisol3}, where for each of the string we have 
\bea\la{x1a11}
&&x_1(\infty)=\int_{r_v}^\infty dr \sqrt{-\frac{Q_1^2 g_{33} g_{rr}}{g_{11}\prt{Q_3^2 g_{11}+Q_1^2 g_{33}+g_{00} g_{11} g_{33}}}}~,\\ \la{x3a33}
&&x_3(\infty)=\int_{r_v}^\infty dr \sqrt{-\frac{Q_3^2 g_{11} g_{rr}}{g_{33}\prt{Q_3^2 g_{11}+Q_1^2 g_{33}+g_{00} g_{11} g_{33}}}}~
\eea
and imposing the stability condition.
When there is no field backreaction $g_{11}=g_{33}$, and the quark distribution is symmetric since $Q_1=Q_3$.

Let us examine if there is a universal relation of the shape of the quark distribution and the anisotropic geometry. Using \eq{xisol1} and \eq{xisol3} we get for strings of the same angle
\be\la{distribution}
\frac{x_3'}{x_1'}=\frac{Q_3 g_{11}}{Q_1 g_{33}}~,
\ee
valid for each string. Let us consider a prolate geometry along the $x_3$ direction, $g_{33}>g_{11}=g_{22}$ and define the monotonically increasing function $A(r)$ as
\be
g_{33}(r)=g_{11}(r) A(r)~, \qquad A(r)\ge 1~,
\ee
expressing the fact that we focus on this section on spaces asymptotically AdS $(A=1)$ and the anisotropy increases as we go deeper in the bulk.

For the quarks sitting at angles $\vphi=\pi/4$ and using the fact that constants can be evaluated at a preferred point we get
\be
Q_1=-\frac{g_{00}g_{11} x_1'}{\sqrt{- g_{00}\prt{g_{rr}+\sqrt{2}/{2}x'^2\prt{g_{11} +g_{33} }}}}\Bigg|_{r=r_v}~,\qquad Q_3=\frac{g_{33}}{g_{11}} Q_1\bigg|_{r=r_v} ~,
\ee
leading to the simple relation
\be\la{x3x1pi4}
\frac{x_3'}{x_1'}=\frac{A(r_v)}{A(r)}\ge 1~.
\ee
Using the definition of the derivative, the continuity of the functions and the initial conditions, it can be shown that $x_3(\infty)_{\vphi=\pi/4}>x_1(\infty)_{\vphi=\pi/4}$ and moreover that the strings always bend towards $x_3$.

Next, to find the shape of the quark distribution we need to compare how the quarks are distributed close to the $x_1(\vphi=\pi/2)$ and  $x_3(\vphi=0)$ axes. The computation gives
\bea\la{qi2}
\vphi=0 \Rightarrow    \begin{cases}
 Q_1=0~,\\
 Q_3=-\frac{g_{00} g_{33}  x'}{\sqrt{- g_{00}\prt{g_{rr}+g_{33} x'^2}}}\bigg|_{r=r_v}
 \end{cases},\quad 
 \vphi=\frac{\pi}{2} \Rightarrow    \begin{cases}
 Q_1=-\frac{g_{00} g_{11}  x' }{\sqrt{- g_{00}\prt{g_{rr}+g_{11} x'^2}}}\bigg|_{r=r_v} ,\\
 Q_3=0~,
 \end{cases}
\eea
to find that
\be\label{q1q3case}
\frac{Q_1^2}{Q_3^2}=\frac{g_{11} \prt{\frac{g_{00} g_{rr} }{g_{33} x'^2}+1}}{g_{33}  \prt{\frac{g_{00} g_{rr}}{g_{11} x'^2}+1} }\Bigg|_{r=r_v}\le 1~,
\ee
where both numerator and denominator are considered negative, meaning that we are looking mostly on the stable region of baryon. Otherwise one has to  consider cases. Using this inequality we end up having
as the only possible choice
\be  \label{x3x10pi2}
\frac{x_1'^2{}_{\vphi=\pi/2}}{x_3'^2{}_{\vphi=0}}>1
\ee
and by applying calculus methods and taking into account the boundary conditions for the functions $x_1$ and $x_3$ we find that $x_1{}_{\vphi=\pi/2}>x_3{}_{\vphi=0}$.

Therefore, in any prolate geometry, i.e. $ g_{33}\ge g_{11}=g_{22}$, with the assumptions made for \eq{q1q3case}, the quark distribution associated to the baryonic vertex with the initial conditions considered, is a squashed cycle along the oblate directions, i.e. $x_1{}_{\vphi=\pi/2}>x_3{}_{\vphi=0}$. In other words by applying a strong field in a theory, which backreacts and deforms the dual geometry making it prolate, the cyclic distribution of the quarks in the bound state is squashed with its long axis being in the opposite direction of the long axis of the deformed geometry.

\subsubsection{Quark Distribution for a Baryon on the Transverse Plane}

When the $N_c$ quarks are distributed on the plane $(x_1,x_2)$ with rotation symmetry, the strings are parametrized by $(x_1(\s),~x_2(\s),x_3=0)$, and the equations \eq{xisol1} for $x_1$ and trivially for $x_2$ are equal to
\be
x_1(\infty)=x_2(\infty)= Q_1 \int_{r_v}^\infty\sqrt{-\frac{g_{rr}}{g_{11}\prt{2 Q_1^2+g_{00}g_{11}}}}~.
\ee
We can express the constants of motion with respect to the geometric features of the configuration.
Combining the stability condition  \eq{forcedp1} and the constants of motion \eq{qi} we get
\be
Q_1^2=-\frac{1}{2}\prt{\frac{g_{11}}{16 g_{rr}}\partial_r \prt{\sqrt{-g_{00}}g_{\th\th}^{5/2} e^{-\phi}}^2+g_{00} g_{11}}\bigg|_{r=r_v}~.
\ee
The integrand depends on the anisotropies through the distortion that cause the external fields to
the geometry. The constant $Q_1$ is a measure on how much the anisotropy affects the radius of the symmetric baryon, and we will demonstrate below its quantitative role in certain theories.

In the last two sections we have developed the formalism to determine the stability condition of the baryon, the energy and the shape.  In the following section we apply our generic methods to certain IIB supergravity solutions dual to field theories with a space-dependent axion generating an anisotropy along one direction.

\section{Baryon in Theories with  Space Dependent $\theta$-Term} \la{section:theta}

\subsection{The Theory}

Let us apply the baryon probe analysis we have developed in the previous sections to the dual of a field theory which is a deformation of the $\cN=4$ sYM with a $\th$-term depending linearly on the spatial dimensions of the space as
\be
S=S_{{\cal N}=4 \mbox{ sYM}} +  S_\th~,\qquad   S_\th   = \frac{1}{8\pi^2} \int   \th(x_3) ~ \mbox{Tr}~ F \wedge F ~,\qquad \th=\frac{8\pi^2 \pi N_c a }{\l} x_3~,
\label{gaugeaction1}
\ee
where $\l$ is the t' Hooft coupling and $a$ is a constant. Our theory has broken isotropy, the CP is broken,  and it is translational invariant, respecting all the global symmetries that are accommodated in the gravity dual theory \eq{metric1}.

The top-down IIB supergravity dual theory in finite temperature has a non-trivial axion and dilaton. The $\th$-parameter is related through the complexified coupling constant to the axion of the gravity dual theory and therefore one expects a linear  $\chi=a x_3$ dependence. The relevant part of the action in the string frame is
\be
S=\frac{1}{2 \kappa_{10}^2}\int d^{10}x \, \sqrt{-g} \prt{e^{-2\phi}(R+4\partial_M\phi \partial^M\phi)
-\frac{1}{2}F_1^2-\frac{1}{4 \cdot 5!}F_5^2
}~ ,
\label{sugraaction}
\ee
where the index $M=0,\ldots,9$ and the field strength of the axion is $F_1=d\chi$. The system of the ordinary differential equations of the above action can be solved numerically.  The axion is the source of anisotropy and the parameter $a$ quantifies its strength.  The Hogde-dual of the axion field strength is the field strength of a RR 8-form $C_8$ on the base $C_{x_0x_1x_2\th^i}(r)$ where $\th^i$ are the coordinates describing the internal sphere $S^5$. Therefore the picture with the axion field, is equivalent to the backreaction of D7-branes extending along $(x_0,x_1,x_2,\th^i)$ on the near horizon limit of the D3 branes, forming the AdS. The anisotropy
is generated due to the D7-brane distribution along the   direction $x_3$. Naturally the anisotropic parameter $a$ can be thought also as the density number of the D7-branes backreacting on the D3-branes.

Then the string frame metric takes the form
\bea \label{anisometric}
&&ds^2 = r^2 \prt{ -\cF(r) \cB(r)~ dx_0^2+dx_1^2+dx_2^2+\cH(r) dx_3^2} +\frac{ dr^2}{r^2 \cF(r)}+ {\cZ(r)} ~ d\Omega^2_{S^5}~ ,\\
&&\cH(r)=e^{-\phi(r)}~,\qquad \cZ(r)=e^{\frac{1}{2}\phi(r)}~,
\eea
where $\cF\propto(r-r_h)$ is the blackening factor with a horizon at $r_h$, $\cH(r)$ is the anisotropic spatial deformation of the geometry along the holographic direction. The solutions of the above action describe an RG flow from an AdS geometry at the boundary, to an anisotropic Lifshitz-like geometry in the IR with fixed scaling exponent $z=3/2$ \cite{Azeyanagi:2009pr,Mateos:2011tv}. The reason that the scaling turns out to be fixed in supergravity theories was explained in \cite{Giataganas:2017koz}, and it is related to the fixed axion-dilaton coupling in the supergravity theories. Relaxing this coupling by allowing a new parameter, Lifshitz-like IR geometries with arbitrary scaling exponents can be obtained \cite{Giataganas:2017koz}.

We find the background solutions numerically and analytically by perturbative techniques in the low $a/T$ region. Since for our purposes the analysis of the quark baryon distribution and the energy  require numerical techniques, irrespective of having the gravity background in an analytic form or not, we decide to work mostly with the numerical full range supergravity solutions. Although this analysis is more demanding it has the advantage  we can go away of the perturbative low anisotropy $a/T$ regime, where the anisotropic effects of the theory are clear on the baryon. Nevertheless we include few semi-analytical results in the next section.

\subsection{Baryon Radius, Shape  and Anisotropy}
\subsubsection{Baryon Radius and Shape: Semi-Analytic Study}

The computations of the quark baryon distribution involves differential equations which are not solvable
analytically in the general case. However, at certain limits  by making approximations we obtain semi-analytical results. In this section we present some of them in a compact notation.

At the low anisotropy, high temperature limit the background \eq{anisometric} is given in Appendix \ref{app:lowat}. The radius of the baryon on the boundary is angle dependent and its value depends on the ratio of the position of the black hole horizon to the position of the D5-brane as well as on the strength of the anisotropy of the theory. Let us introduce  the dimensionless quantities $\rho:= r_h/r_v$ and $\rt:=r/r_v$, where $r_v$ is the position of the D5-brane. The net force on the baryon can be found by applying \eq{forcedp2} on \eq{anisometric} and the lengthy expansion is written in a schematic form as
\be
F=F_0(\rho,r_h,x'_v)+a^2 F_2(\rho,r_h,x'_v,\phi)~,
\ee
where $F_0$ is independent of the angle as expected, in contrast to the anisotropy contributed term $F_2$ signaling the dependence of the quark distribution on the angle. The $x'_v:=x'(\s_v)$ is the derivative of the string at the brane, describing the angle that is attached to it. By integrating the above equation on all angles in the continuum limit, and requiring stability we obtain\footnote{The net force is an even function and there are two solutions of $x'_v$. The final outcome is independent of the sign of the solution we choose as long as the signs are kept in all equations consistent. Here we choose positive $x'_v$.}
\be \la{solxb1}
x'_v=x_{v0}'(\rho,r_h)+a^2 x_{v2}' (\rho,r_h)~.
\ee
To find the effect on the shape of the distribution we have to compute the string along $x_1(s)$ and $x_3(s)$. By substituting the constants, the string differential equations \eq{x1a11} and \eq{x3a33} for fixed angles eventually depend on metric, and $x'_v$. One can obtain the lengthy relations of the form
\be
x_{1,\vphi=\pi/2}=\int_{1}^{\infty} d\rt (x_{1}^{(0)}+a^2 x_{1}^{(2)})~,\qquad x_{3,\vphi=0}=\int_{1}^{\infty} d\rt (x_{3 }^{(0)}+a^2 x_{3 }^{(2)})~,
\ee
where $x_{1 }^{(0)}= x_{3 }^{(0)}$, since the axion affects the 2nd order expansion, to give:
\bea\label{seca2}
x_{1,\vphi=\pi/2}^{(2)}-x_{3,\vphi=0}^{(2)}&=&  \int_{1}^{\infty} d\rt X_{13}\prt{x'_b, \rho,\rt}~,
\eea
where the function $X_{13}$ is given explicitly in the equation \eq{x13lowat} of the Appendix \ref{app:lowat}. In the low anisotropy limit, this is the anisotropy contribution to the deformation of the cyclic  quark distribution given by the 0th order terms. Notice that the zero anisotropy limit in our analysis is smooth.

\subsubsection{Baryon Radius and Shape:  Full Range Study}

Let us use the generic analysis of the section \ref{section:generic} and apply it to the axion-dilaton supergravity theory. We firstly solve numerically the equations obtained by the action \eq{sugraaction} to generate the background \eq{anisometric}. In the parametric region we work we obtain prolate dual geometries with $g_{33}>g_{11}=g_{22}$. As a side remark we point out that the pressure anisotropy of the theory does not follow necessary the above geometric inequality.

We follow certain steps on the numerical study for the baryon in anisotropic plane which we spell out below. Having obtained the gravity dual theory we solve the net force condition \eq{forcedp2} for a chosen particular position of the D5-baryon brane. Therefore we can determine the value of the derivative of the string hitting the brane such that the system has zero force. Next, we evaluate the constants of motion $Q_1$ and $Q_3$ using the equations \eq{qi}. We choose to compute them at the radial position of the brane, taking advantage of their constant nature. Then to determine the radius of the distribution along different directions we integrate the equations \eq{xisol1} and \eq{xisol3}. We observe that the distance between the antipodal quarks  of the baryon depends on the angular position of the quarks. Therefore the distribution of the $N_c$ quarks is not a cyclic one. The radius of the distribution is smaller along the anisotropic direction, counterbalancing the anisotropic geometry deformation in agreement with the generic results of section \ref{section:shape}.

\begin{figure}
\begin{minipage}[ht]{0.5\textwidth}
\begin{flushleft}
\centerline{\includegraphics[width=80mm]{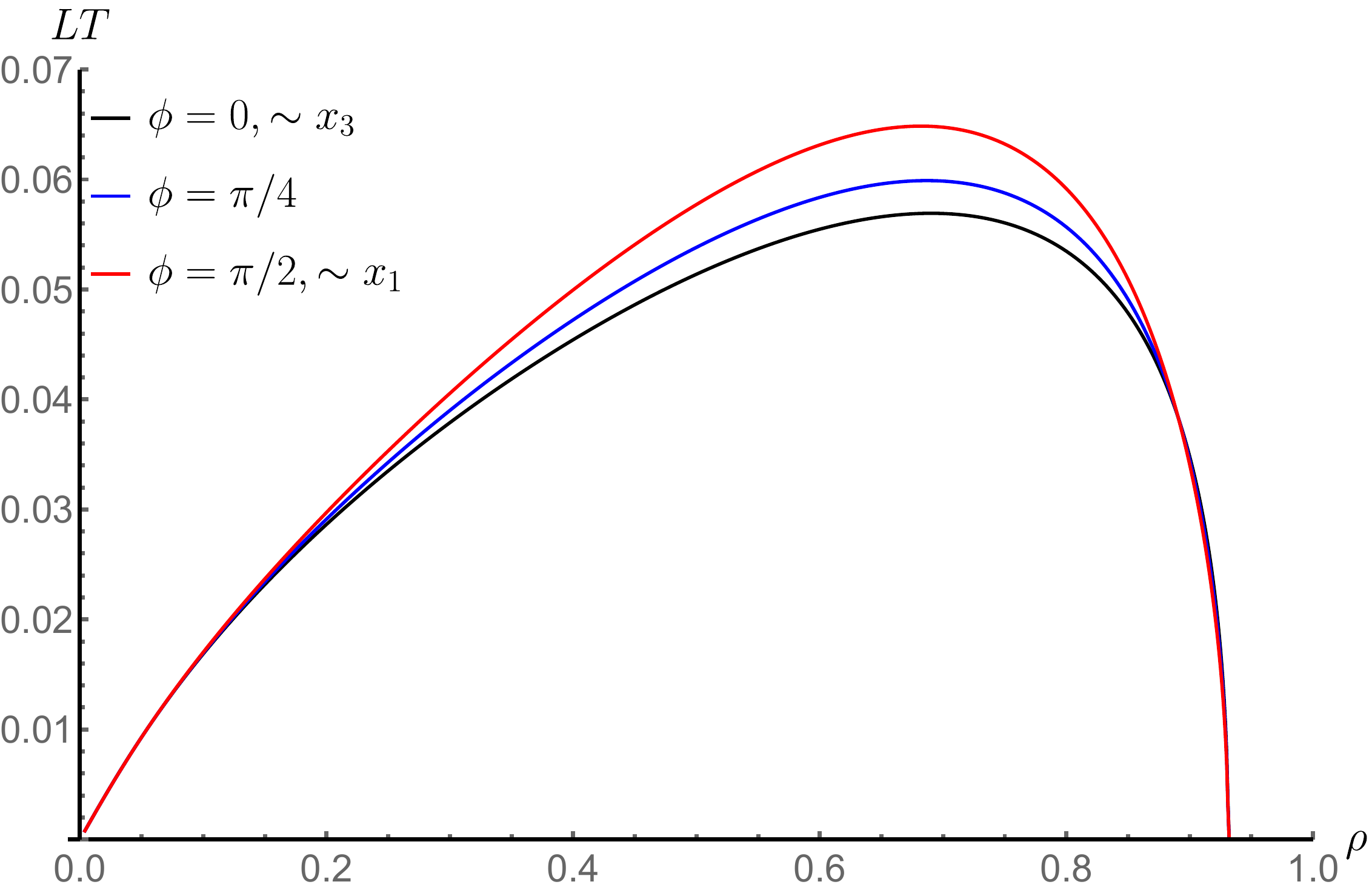}}
\caption{\small{The radius of the baryon vs the brane vertex position for different angles when the quarks are distributed on $\prt{x_1,~x_3}$ plane and for fixed anisotropy. Notice that transverse to the anisotropy, the distribution is affected weaker and the maximum of LT is larger hinting that this the most stable direction as we increase the temperature.}}
\label{fig:ltrho1}\vspace{.6cm}
\end{flushleft}
\end{minipage}
\hspace{0.2cm}
\begin{minipage}[ht]{0.5\textwidth}
\begin{flushleft}
\centerline{\includegraphics[width=80mm ]{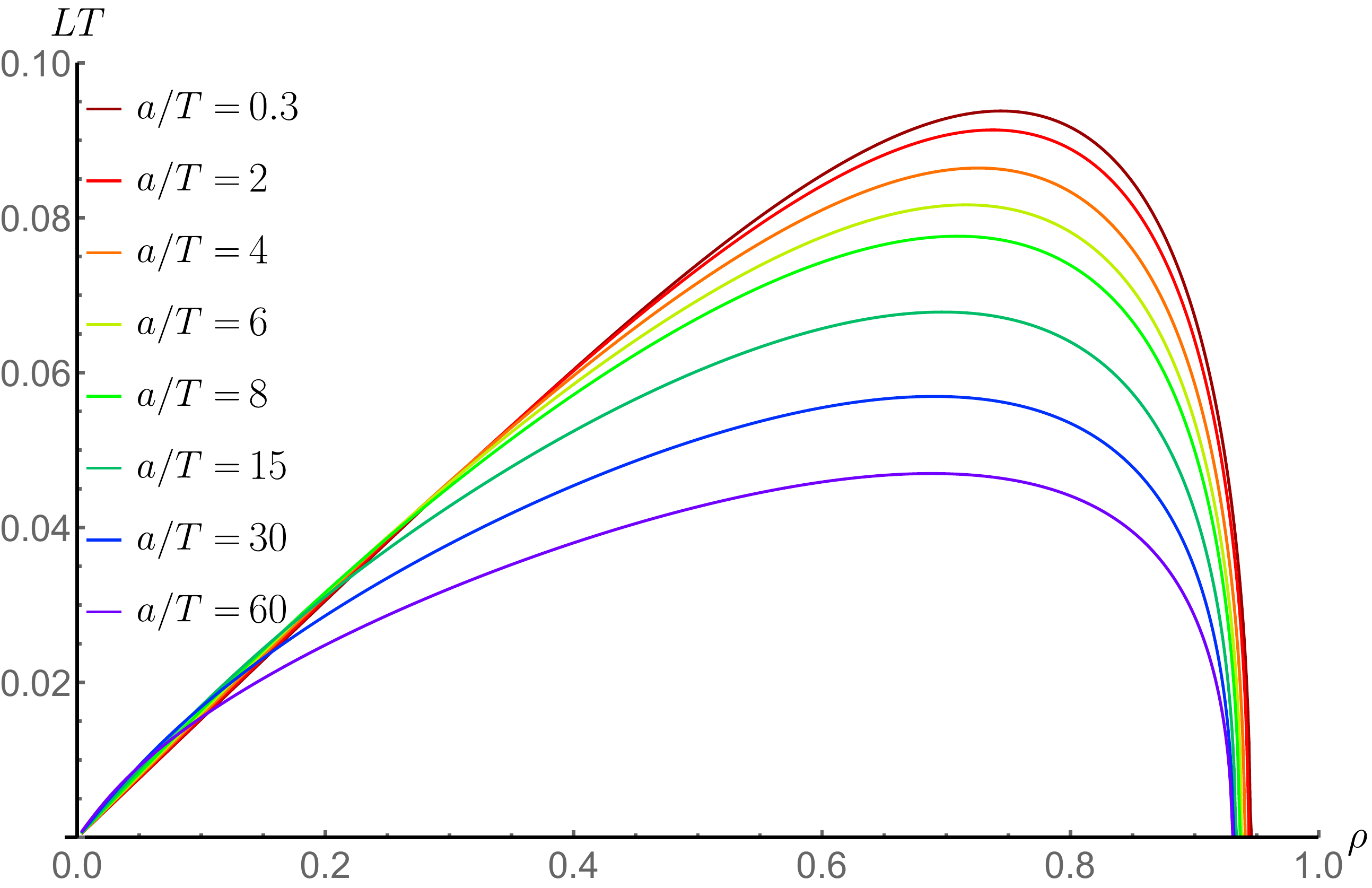}}
\caption{\small{The radius of the baryon along the anisotropic direction $x_3$ ($\phi=0$) for different values of the anisotropy. Increase of the anisotropy results to more severe deformation of the distribution and to easier dissociation of the bound baryon state i.e. lower critical temperature. A rainbow coloring has used for the plot with respect to the value of the anisotropy.} }
\label{fig:ltrho2}\vspace{.cm}
\end{flushleft}
\end{minipage}
\end{figure}

Implementing the above steps we present in Figure \ref{fig:ltrho1} the radius of the distribution for three different angles while we vary the position of the vertex brane. Quarks distributed transverse to the anisotropic direction are affected the less. Moreover we observe that the stability of the baryon becomes more complicated than the usual cases. The isotropic baryon becomes unstable beyond the maximum point of the function $LT(\r)$, corresponding to a natural phase transition and competition for dominance between two different energy branches related to the two possible baryon configurations. In the anisotropic case the maximum point depends on the angle with respect to the anisotropic direction. A natural explanation is that the baryon starts to become unstable at the lowest value satisfying $LT'(\r)=0$  for any corresponding angle $\phi$. This suggests that the baryon  becomes unstable in discrete stages with respect to the position of the quarks forming it.  In an anisotropic $SU(N_c)$ theory, the quarks along the anisotropic directions will become unstable first, leaving a baryon with $N_c-2$ quarks and this sequence will continue with the pair of quarks closer the anisotropy until the baryon becomes an unstable state.

Moreover, we find that increase of the number of the backreacting anisotropic $D7$-branes leads to further increase of the ellipsoid like distribution of the quarks on the boundary and   decrease of the baryon radius. This is naturally expected,
by increasing the strength of the axion field we increase  the anisotropic gluon interactions, which affect more the quark distribution. The relevant results are presented in Figure \ref{fig:ltrho2}.

\subsection{Quark Distribution of the Boundary State and Anisotropy}

\subsubsection{Quark Distribution on the Anisotropic Plane}
The full shape of the baryon configuration is obtained by computing the string profiles along $x_1(r)$ and $x_3(r)$. The projection of the strings on the spatial plane shows the degree of the squashing of the baryon compared to the squashing of the anisotropic spacetime. Increase of the anisotropy deforms the cyclic distribution and at the same time results to increased deformation of the strings' straightness. The analysis consists of solving explicitly the string equations \eq{x1a11} and \eq{x3a33} for each angle $\vphi$ to determine the full shape of the strings. To do that we have to firstly repeat the numerical steps described in the previous section, solving the zero force condition in order to determine the $x'_v$ and the constants $Q_i$.

\begin{figure}[ht]
\begin{minipage}[ht]{0.5\textwidth}
\begin{flushleft}
\centerline{\includegraphics[width=80mm]{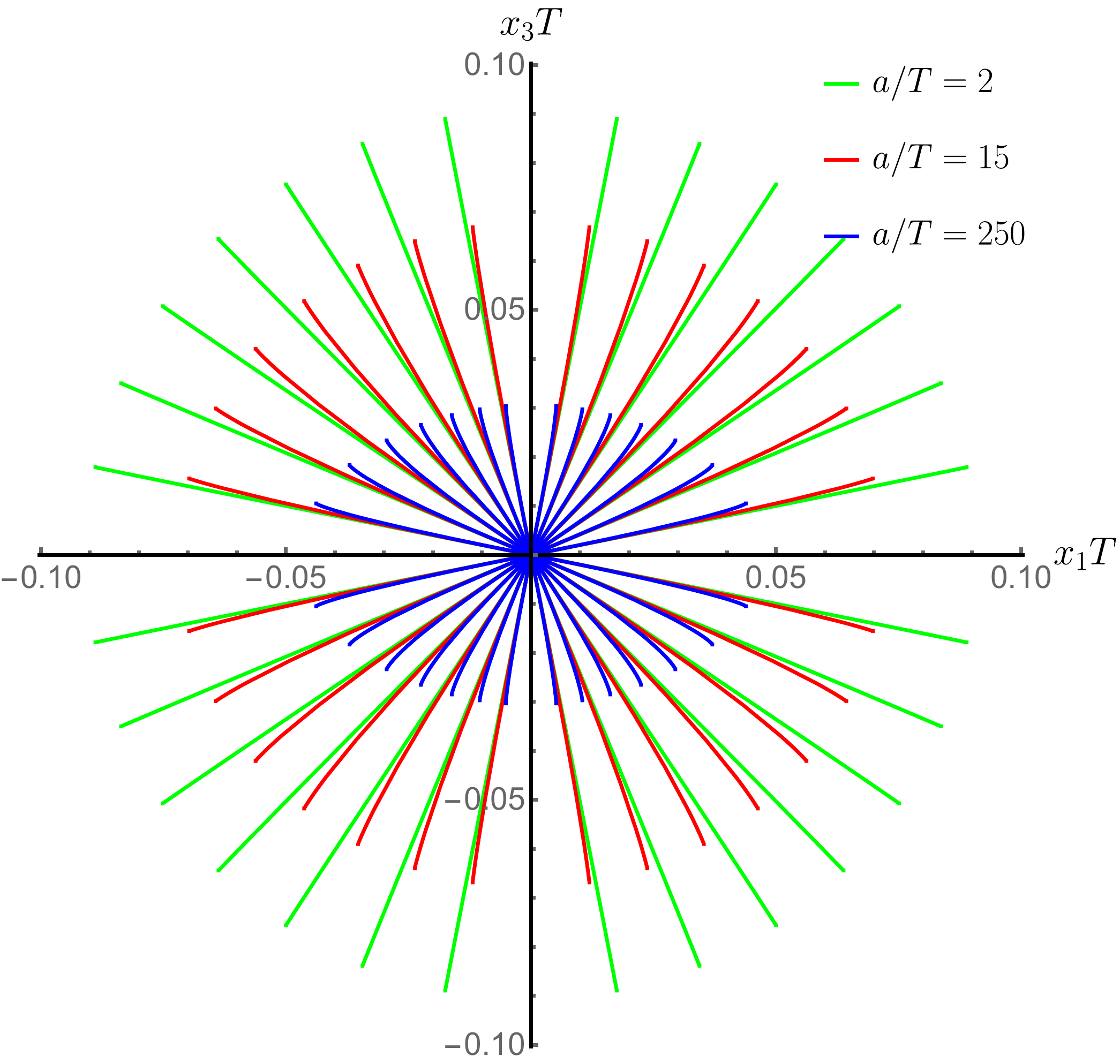}}
\caption{\small{The strings of the baryon projected in the $\prt{x_1,x_3}$ plane. The strings meet at the center where the D5-brane sits in the bulk, while the endpoints of the strings terminate at the UV boundary, where the quarks are located. Notice that the shape of the distribution becomes increasingly elliptic as the anisotropy increases with long axis $x_1$ as described by \eq{x3x10pi2}. Increase of the anisotropy leads to bending of the strings towards $x_3$ as described by \eq{x3x1pi4} and to lower radii for the quark distribution at the boundary. The brane location in the bulk is kept constant.}}
\label{fig:ltdistribution}\vspace{.0cm}
\end{flushleft}
\end{minipage}
\hspace{0.3cm}
\begin{minipage}[ht]{0.5\textwidth}
\begin{flushleft}
\centerline{\includegraphics[width=47mm ]{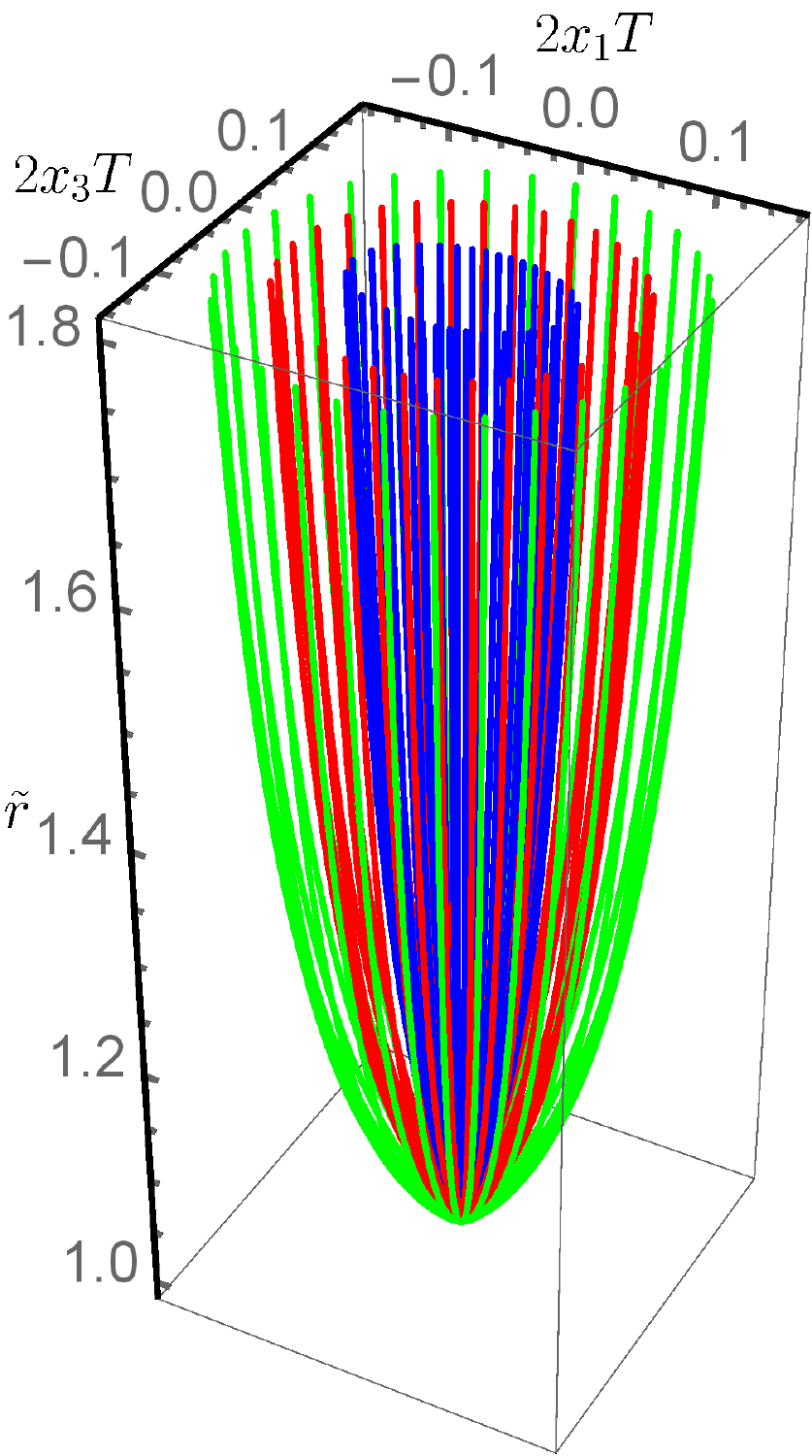}}
\caption{\small{The baryon vertex as computed in the holographic space for $a/T=2,15,250$ (green, red, blue), where increasing values of anisotropy lead to smaller boundary radius. The endpoint of the ''cigar" where the strings meet is the location of the vertex D5-brane in the bulk of the space. The strings terminate at the boundary where the quarks are located. Notice the deformed circle of the inner quark distribution  where the effects of anisotropy on the baryon are stronger. The spatial scales are magnified by a factor of two for presentation reasons. The holographic direction is normalized with the Dp-brane position $\tilde{r}=r/r_v$.} }
\label{fig:dist3d}\vspace{.0cm}
\end{flushleft}
\end{minipage}
\end{figure}

The study gives a nice graphic realization of the projection of the fundamental strings on the $(x_1,x_3)$ plane in Figure \ref{fig:ltdistribution}. It is remarkable how the shape of the baryon is modified with increasing anisotropy in order to retain its stability. Notice the increasing deformation of the distribution  towards an elliptic-like shape and the decrease of its radius. The bending of the strings is also obvious and is more severe for mid-quarter strings.

Then we obtain numerically the full baryon-string vertex and we depict our solution in the holographic space in Figure \ref{fig:dist3d}. The shape of the configuration is of cigar-type and  depends on the value of anisotropy. Notice that the radial distance of the vertex is kept fixed and the strings become relatively quickly transverse to the boundary as they depart from the D5-brane.

\subsubsection{Quark Distribution on the Isotropic Plane}

\begin{figure}
\begin{minipage}[ht]{0.5\textwidth}
\begin{flushleft}
\centerline{\includegraphics[width=80mm ]{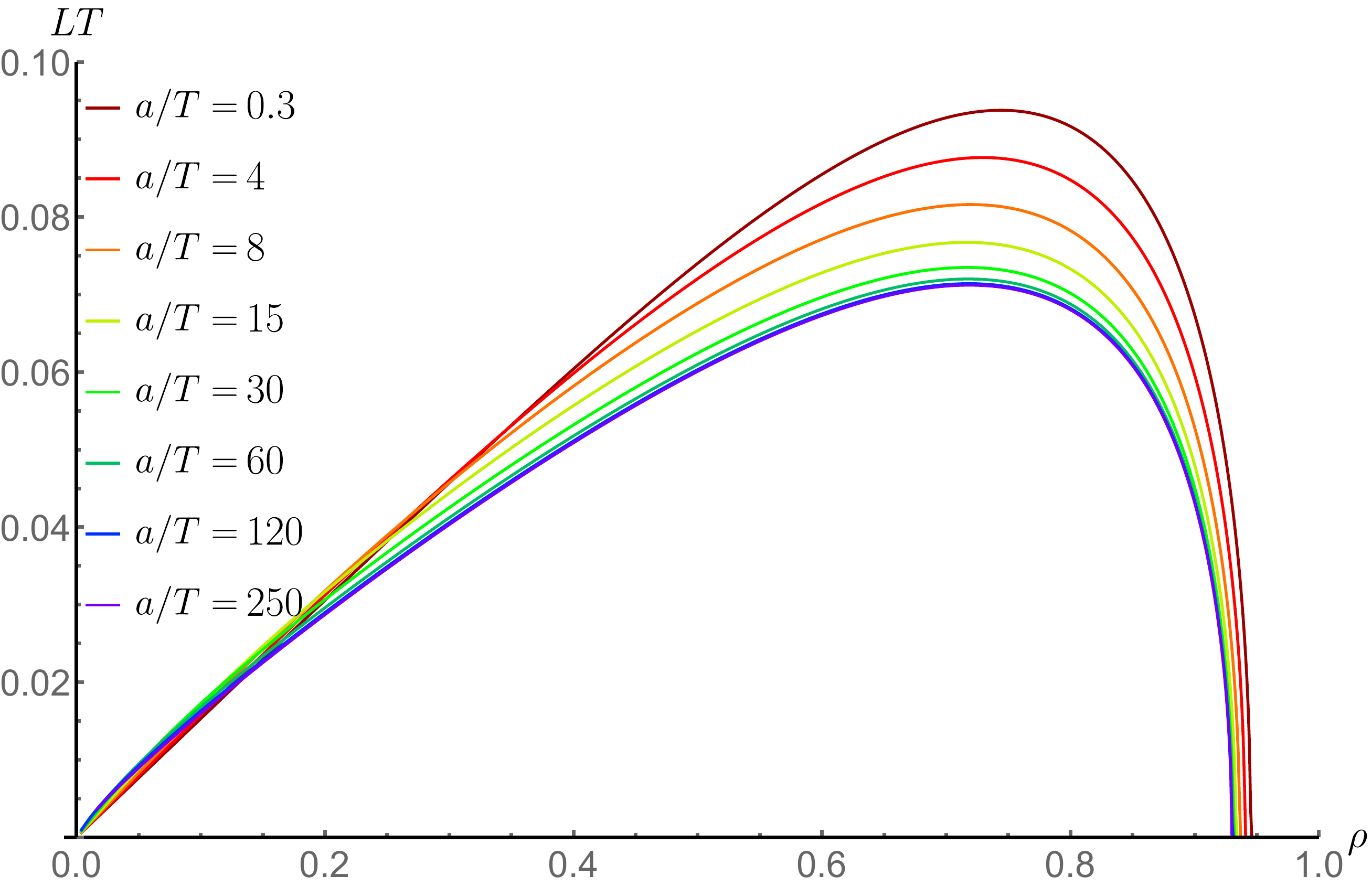}}
\caption{\small{The radius of the baryon   on the rotational invariant $(x_1,x_2)$ plane, for different values of the anisotropy. Notice the decrease of the radius with the strength of the field and that for large strength, the baryon has much weaker dependence on it. This is depicted in the plot with the increasing density of curves.} }
\label{fig:traltrho2}\vspace{-1.5cm}
\end{flushleft}
\end{minipage}
\hspace{0.3cm}
\begin{minipage}[ht]{0.5\textwidth}
\begin{flushleft}
\centerline{\includegraphics[width=70mm]{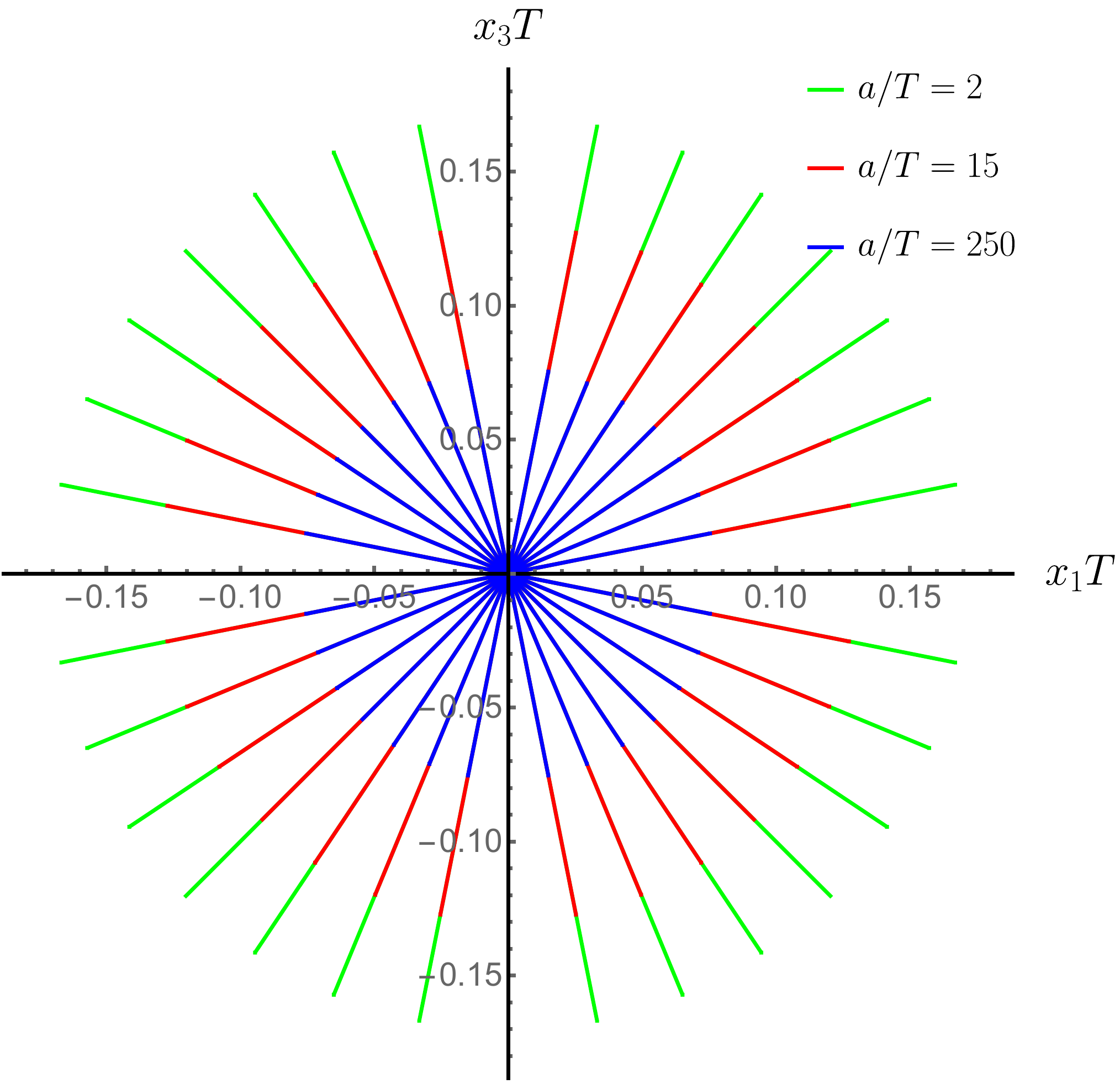}}
\caption{\small{The shape of the fundamental strings projected on the $(x_1,x_2)$ plane. Each string terminates at the boundary, while their meeting point is where the vertex brane is located.  The reduction of the baryon size and the completely circular distribution is observed, in contrast to the distribution on the anisotropic $(x_1,x_3)$ plane of Figure \ref{fig:ltdistribution}.}}
\label{fig:tra2d}\vspace{0cm}
\end{flushleft}
\end{minipage}
\end{figure}

Here we repeat the numerical analysis for a baryon where its $N_c$ quarks lie on the rotational invariant $(x_1,x_2)$- isotropic plane. The properties of the baryon are still affected by the external field creating the anisotropy since the gluons interact with the bound state in a different way when fields are present. By applying the analysis of the section \ref{section:generic}, we compute the radius of the baryon and we find that increase of field strength leads to decreasing baryon radius and easier dissociation. In this case several simplifications occur, since the strings along $x_1$ and $x_2$ are equivalent due to rotational invariance in these directions \footnote{Notice that for the strings of this section we have $(x_1(\s),~ x_2(\s),~x_3=0)$.}. The constants $Q_i$ at the equations \eq{qi} satisfy  $Q_1=Q_2$, while the equations of motion \eq{x1a11}
give
\be\la{xitra}
x_1'^2(s)=x_2'^2(s)=-\frac{Q_1^2 g_{rr}}{ g_{11}\prt{2Q_1^2 +g_{00} g_{11} }}~.
\ee
The ensure  stability of the state along the holographic direction we still get the same equation  \eq{forcedp1}, where $\cD_1$ contains the $x_1$ and $x_2$ directions.  The numerical treatment goes as in the previous section and the results are presented in Figure \ref{fig:traltrho2}, where observe a decrease of the radius with the strength of the field.

Since the baryon lies on the transverse rotational invariant plane we expect that it will be symmetric with respect to the origin and to have straight strings solutions in contrast to the previous setting. This is due to having two equivalent differential equations given by \eq{xitra} and when solving them we obtain the shape of the full string depicted in Figure \ref{fig:tra2d}. Notice that in order to retain the state stable, the quarks organize in such a way to come closer to their antipodal partner as the anisotropy increases. Moreover, notice the weaker dependence of the isotropic baryon on the increase of anisotropy compared to the baryon located on the anisotropic plane.

\subsection{The Energy and the Dissociation of the Baryon}

We have observed that the fundamental strings from the baryon's vertex point of view are not all equivalent to each other when the baryon lives on the anisotropic plane. To compute the total energy of the baryon one has to compute the regularized energy of each string and the energy of the baryonic D5 brane vertex. The strings in the direction parallel to the anisotropy are mostly affected by it and are the ones that are pulled stronger closer to the D5-brane and therefore are expected to be the ones with lowest dissociation length $L_c$.

Applying the formalism of the section \ref{section:generic} to the theory with the space dependent $\theta$-term \eq{anisometric}, we compute  the energy of a baryon  in the theory.
We firstly solve the differential equations  \eq{x1a11} and \eq{x3a33}, to obtain the radius of the state and then we integrate \eq{enea1} to get the energy. We present here directly our numerical results.

\begin{figure}[ht]
\begin{minipage}[ht]{0.5\textwidth}
\begin{flushleft}
\centerline{\includegraphics[width=84mm ]{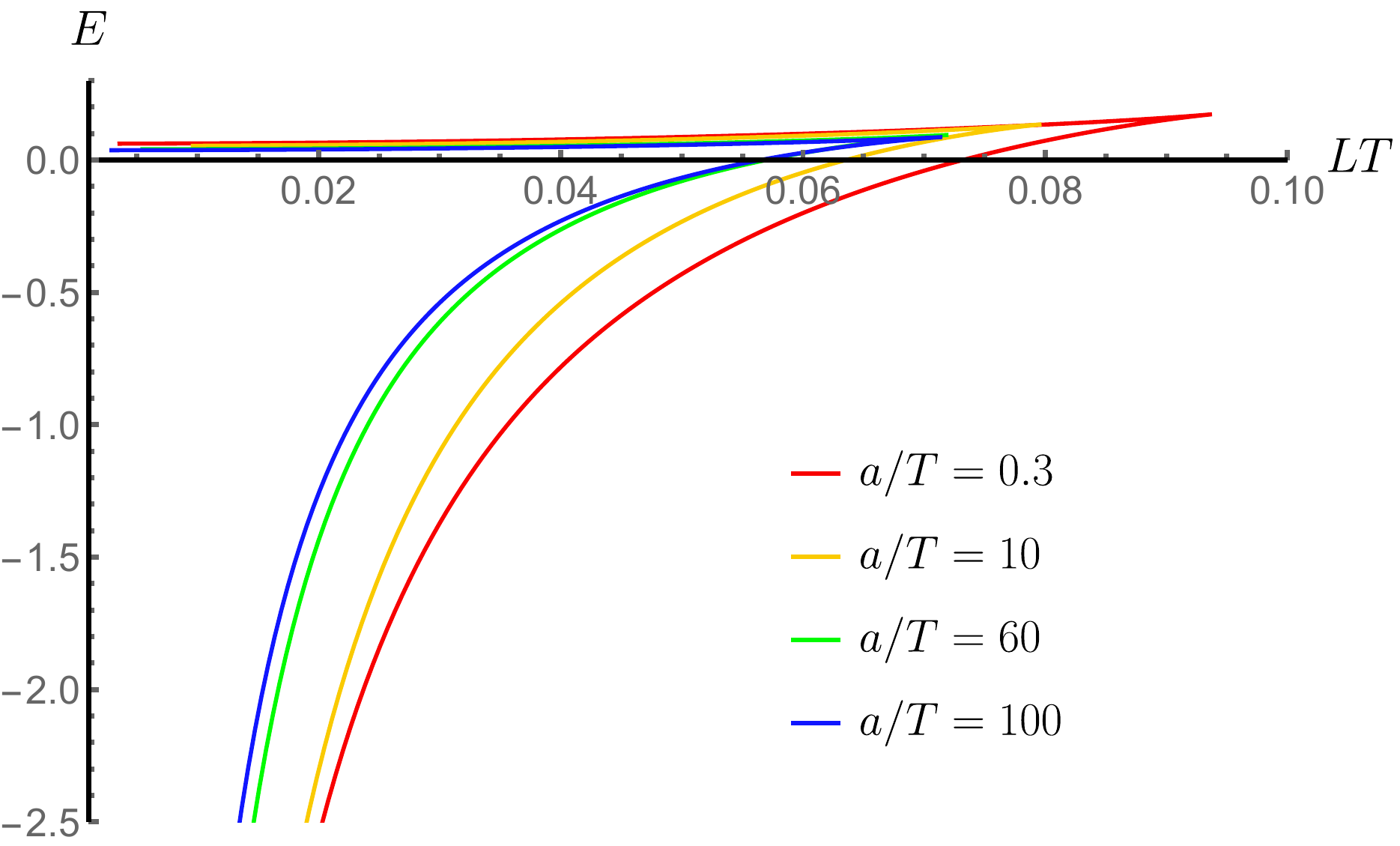}}
\caption{\small{The energy of the baryon vertex on the transverse plane with rotational symmetry. For increasing strength of anisotropy, the decrease of the dissociation length is clear. Equivalently, the increase of the anisotropy leads to increase of the baryon energy of fixed size leading to easier dissociation. Notice that for large values of anisotropy, variations of the anisotropy strength do  not affect the baryon significantly, where the inner energy curves the baryon on the plot approach to a single one. } }
\label{fig:traelt}\vspace{2.1cm}
\end{flushleft}
\end{minipage}
\hspace{0.3cm}
\begin{minipage}[ht]{0.5\textwidth}
\begin{flushleft}
\centerline{\includegraphics[width=85mm]{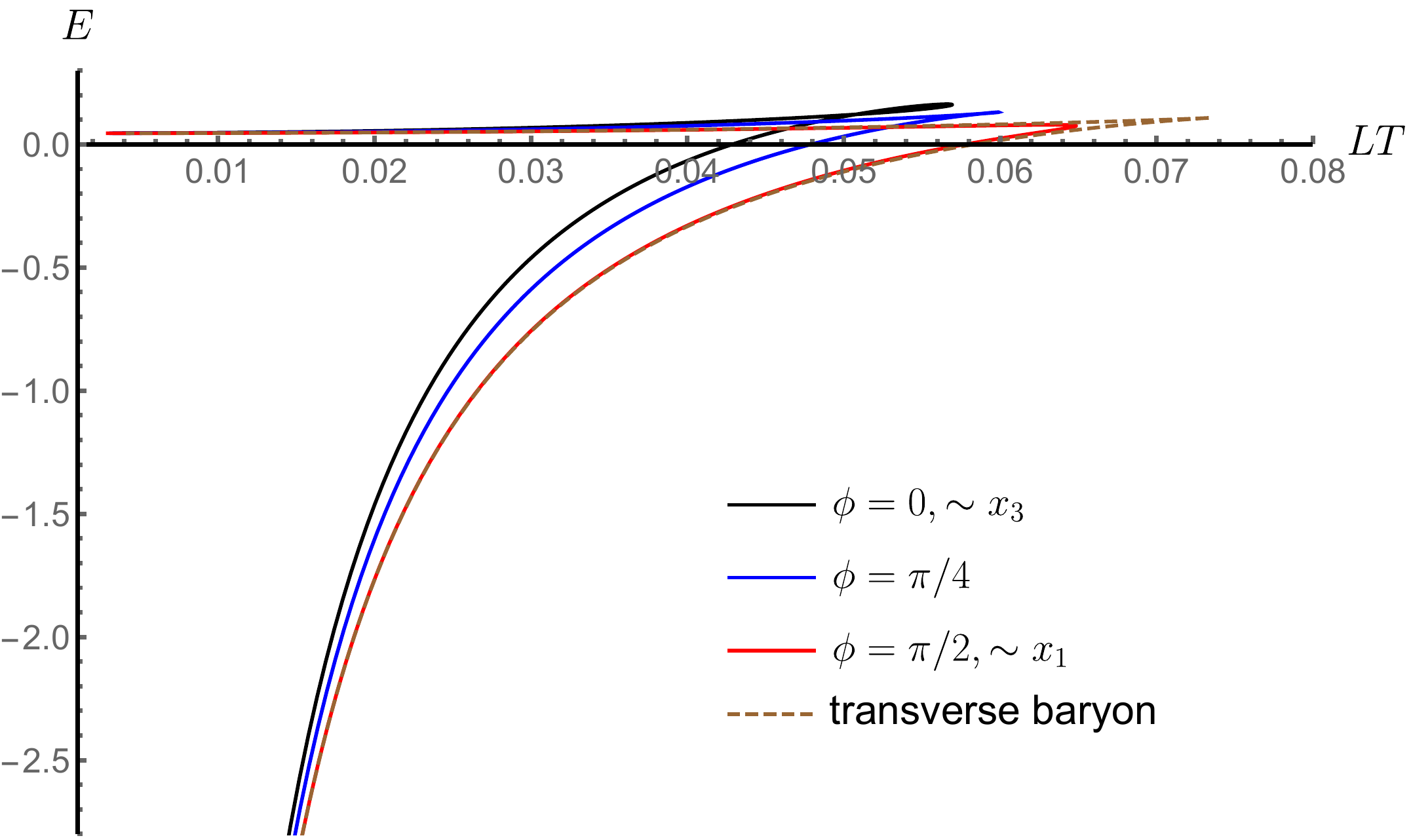}}
\caption{\small{The comparison of the energy of the baryon vertex with quarks distributed on the anisotropic plane versus the baryon with quarks on the transverse plane for fixed value of anisotropy $a/T=30$. The baryon on the transverse to anisotropy plane is more stable than the one in the anisotropic plane. Rotating the baryon from $\prt{x_1,x_2}$ to $\prt{x_1,x_3}$ plane the bound state is affected more severe by the anisotropy. Extremal baryons where the quarks are aligned only along transverse and longitudinal directions are presented. The baryon with quarks uniformly distributed on the anisotropic plane will dissociate in stages with the pair of quarks closer to the anisotropic direction leaving the state first.}}
\label{fig:traelt2}\vspace{-.5cm}
\end{flushleft}
\end{minipage}
\end{figure}

By fixing the strength of the field  responsible for the anisotropy we compute the energy of baryonic state,  on  isotropic and the anisotropic plane. The computation is useful to recognise several properties of the baryonic state.   The increase of anisotropy, makes the transverse baryon to dissociate easier since its energy becomes zero for lower baryon radius \footnote{This tendency is in agreement with the easier dissociation of the mesonic bound states observed in \cite{Giataganas:2012zy} and   other relevant studies, for example \cite{Chernicoff:2012bu,Giataganas:2013lga,Fadafan:2013bva,Avila:2016mno}.}.  For high values of anisotropies the holographic state depends weakly on further anisotropy variations (Figure \ref{fig:traelt}), and it seems that there exist certain saturation effects.   The regularized energy of the strings corresponding to
the first bracket of \eq{enea1}, as well as the brane energy corresponding to the second bracket of \eq{enea1} get both decreased with  $a/T$. In the large anisotropy regime it happens that the remaining energy of the single string decreases in absolute value with anisotropy too, in such a way to cancel out the decrease of the first two terms. Partly responsible for the constant asymptotics of the energy at large anisotropies is that the size of the baryon with respect to the position of the brane parametrized by $L(\rho)$ is weakly depending on the anisotropy at this region as seen in  Figure \ref{fig:traltrho2}. This is an interesting observation which we elaborate later in detail.

The holographic state on the $(x_1,x_3)$ anisotropic plane, has energy that depends on the angle of the quarks on the plane with respect to anisotropy. For example, let us consider the extremal example of having a state with all its quarks on the $x_1$ direction. This state needs larger radius or temperature to dissociate where $E(L_c)=0$, compared to the state having its quarks on the $x_3$ direction (Figure \ref{fig:traelt2}). Therefore, a baryon with quarks uniformly distributed on the anisotropic plane will  dissociate in discrete stages with respect to the position of the quarks forming it. In an anisotropic $SU(N_c)$ theory, the quarks along the anisotropic directions will dissociate first, leaving a baryon with $N_c-2$ quarks and this sequence will continue with the pair of quarks closer the anisotropy until the baryon dissociates completely. This is formally justified since beyond the critical length $L_c$ the energy of a baryon with $N_c-2$ quarks, two straight strings from the Dp vertex brane to the black hole horizon plus two infinite straight strings corresponding to the two quarks left the state, in the anisotropic theory is always  less than the  baryon with $N_c$ strings. This phenomenon provides a unique feature of the effects of anisotropies on the baryonic states. It slightly resembles the phenomena noticed on the baryons moving in a hot plasma wind \cite{Athanasiou:2008pz}. Moreover, similar pattern in a different context has been observed in isotropic cubic lattice simulations for the so-called $k$-strings \cite{Pepe:2009in}.

\subsection{Strong Anisotropy and Baryons}

In this section we focus on the study of the stability range and the dissociation length of the baryon. Our results so far hint that there is a saturation on the effect of the anisotropy to the holographic baryon for large strengths of anisotropy and strong fields.

Let us compute the dependence of the maximum radius of stability  $L_m$, on the anisotropy. We define as the maximum radius of stability the maximum of the $LT(\r)$ function or equivalently where the cusp of the function $E(LT)$ appears. The maximum stability radius is not of much physical importance, since it occurs after the dissociation radius. However, it is still a mathematical property of the holographic baryon solution which we examine. Furthermore, we numerically compute the more physical dissociation radius of the holographic baryon $L_c$  where the regularized energy of the baryon is zero,  in relation with the anisotropy of the space.

\begin{figure}
\begin{minipage}[ht]{0.5\textwidth}
\begin{flushleft}
\centerline{\includegraphics[width=83mm]{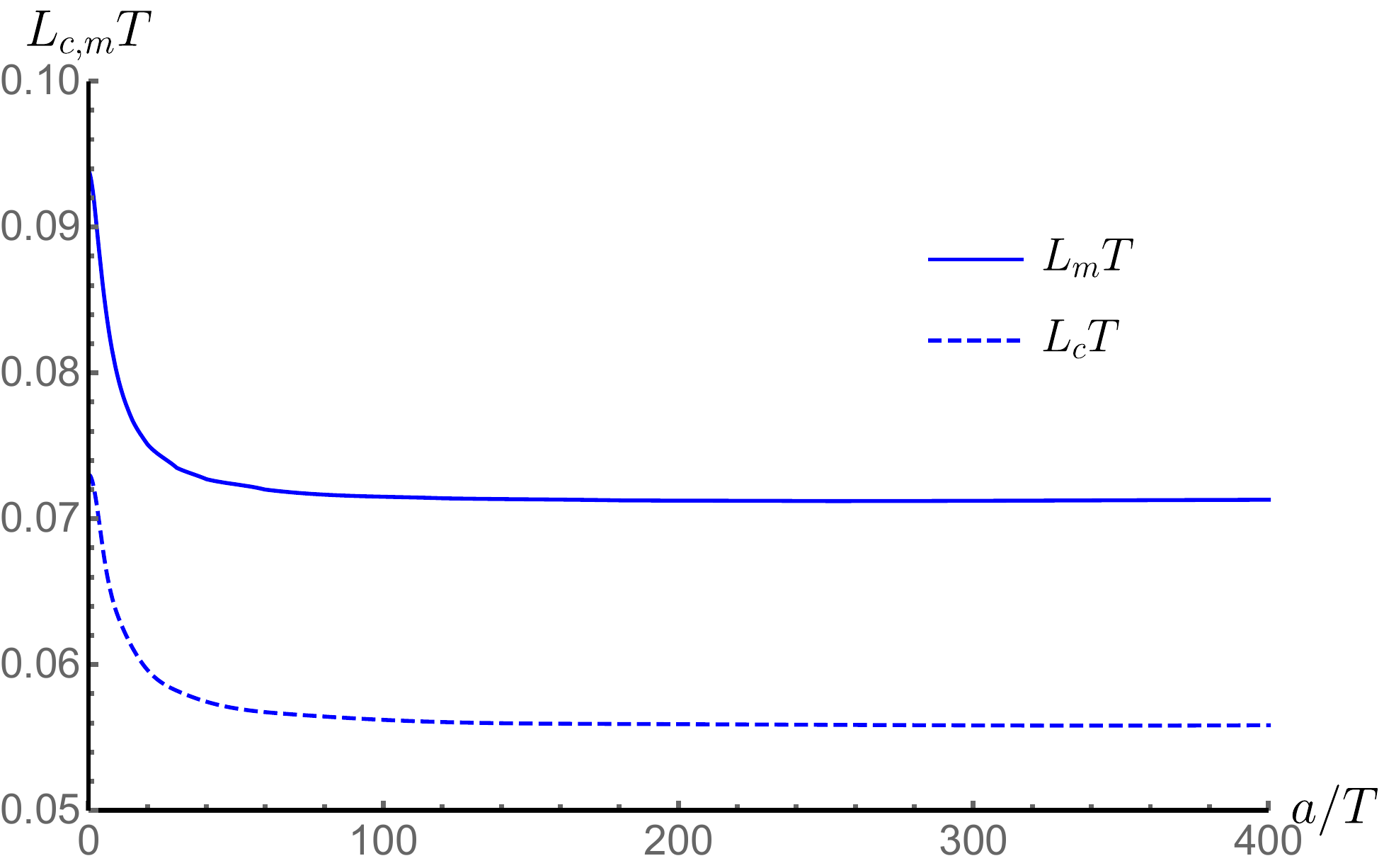}}
\caption{\small{The maximum of the $L(\rho) T$, marked as $L_m$, signals the unstable region which seems to saturate to a value for increasing anisotropy. The dissociation length $L_c$ where the regularized energy of the baryon is zero, also saturates for increasing anisotropy. The numerics are for the baryon located on the isotropic plane.}}
\label{fig:lcmtra}
\vspace{1cm}
\end{flushleft}
\end{minipage}
\hspace{0.3cm}
\begin{minipage}[ht]{0.5\textwidth}
\begin{flushleft}
\centerline{\includegraphics[width=83mm]{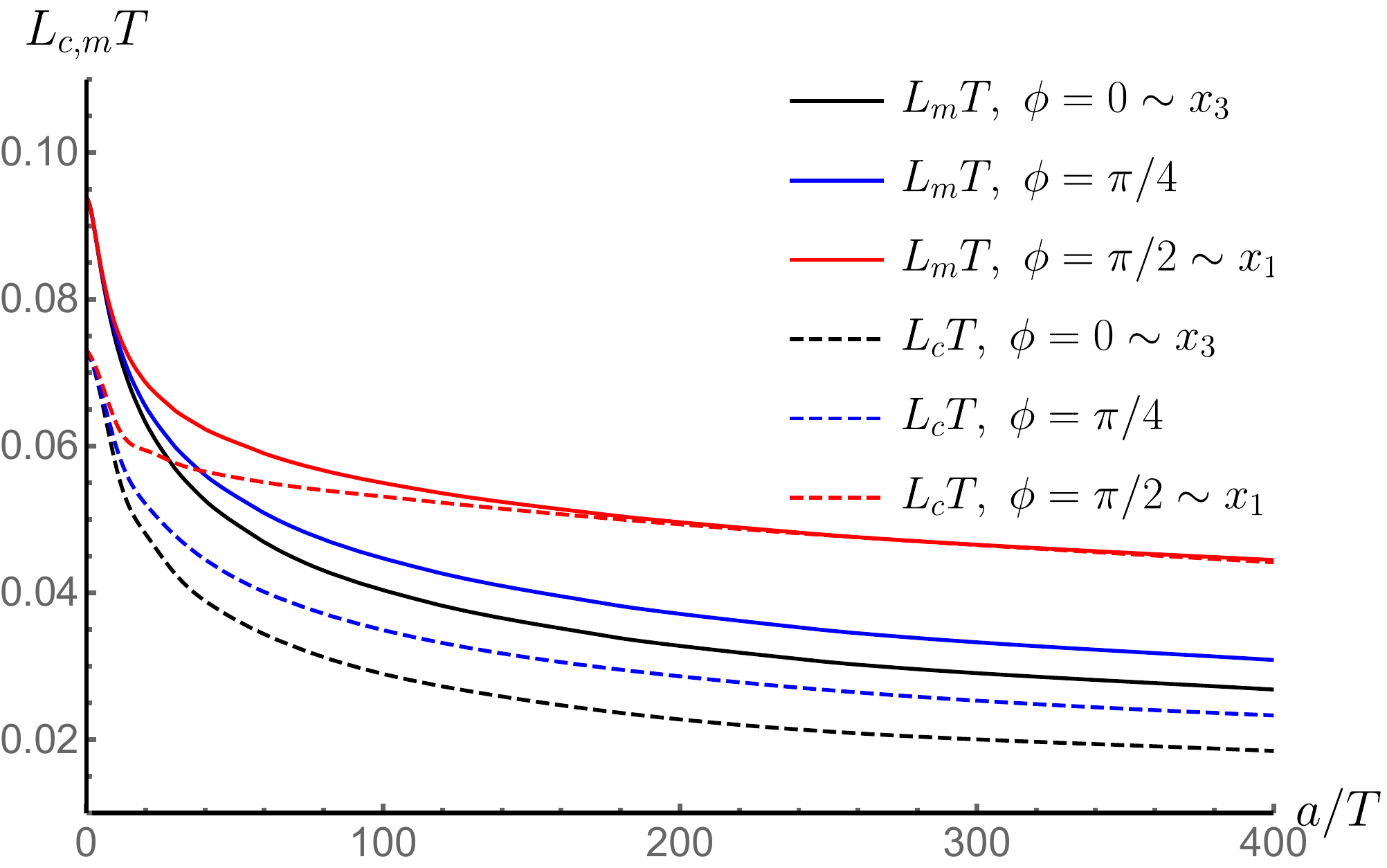}}
\caption{\small{The maximum stability radius $L_m$ and the dissociation length $L_c$ for extremal cases of baryon on the transverse plane, having its quarks on different angles with respect to anisotropic direction. The increase of the anisotropy leads to easier dissociation of the baryon, while for large anisotropy the dependence of the baryon properties becomes weaker on the anisotropic value. The effect is not as severe as in the case of transverse baryon. }}
\label{fig:lcmtpar}
\end{flushleft}
\end{minipage}
\end{figure}

We firstly perform the numerical analysis for the transverse baryon as shown in Figure \ref{fig:lcmtra}.  As we increase the anisotropy, the baryon located transverse to the anisotropic direction, very quickly becomes weakly dependent on the anisotropy. After a certain strength of the external field, the baryon is insensitive to further variations of it signaling an interesting pattern.

The computation is more involved for a baryon being located on the anisotropic plane (Figure \ref{fig:lcmtpar}). Increase of the field strength leads  to weaker dependence of the properties of the baryon on the field's value.   The critical and maximum radius of stability are computed for extremal baryons having their quarks along parallel, transverse and at an angle $\pi/4$ with respect to anisotropy. An interesting observation is that for such baryons the radii $L_m$ and $L_c$ are closer, while in the transverse direction the two radii approach since the $L_m$ is very close to the zero energy.

For large enough anisotropy for a baryon located transverse to the anisotropic direction we propose that the following formula holds
\be
L_c T=c+\cO\prt{\frac{T}{a}}~,\qquad \frac{a}{T}\gg 1~,
\ee
where $c$ is a constant. For a baryon on the anisotropic space, the same formula may hold  with the subleading terms been less suppressed.

In summary, our numerical analysis of this subsection shows that for high enough anisotropy, the baryon tends to depend weaker on it. For the transverse baryon we find that the  dependence becomes negligible, signaling strong saturation effects of anisotropic field on the baryon properties. For the anisotropic baryon, we confirm that irrespective of how large the anisotropy is, the baryon will dissociate in  stages, where the quarks along the anisotropic direction will abandon the state first.

\subsection{High Anisotropy Regime and Baryon in the Exact IR Geometry }

The radii $L_c$ and $L_m$ determined by integrations on whole range of the holographic RG flow. For high enough anisotropy our numerical analysis suggest that the combination of the formulas involved, may pick contributions from a certain holographic region, where the anisotropy changes are suppressed. The baryon analysis depends heavily on the net force conditions, so let us look at the net forces $F_{\parallel}$  and $F_{\perp}$ for quarks located on the plane that includes the anisotropic direction  and for quarks distributed on the $SO(2)$ plane respectively. They take the form
\bea\nn
&&F_{\parallel}=F_{1\parallel}+F_2~,\quad F_{\perp}=F_{1\perp}+F_2~,\quad 16 \pi F_2=\frac{e^{-\phi} g_{\th\th}^4\prt{5 g_{00} g_{\th\th}'+g_{\th\th}\prt{g_{00}'-2 g_{00}\phi'}}}{\sqrt{-g_{00} g_{\th\th}^5}}~, \\\nn
&&16 \pi F_{1\parallel}=  -\frac{8 g_{00} g_{rr}}{\sqrt{-g_{00}\prt{g_{rr}+x'^2 g_{33}}}}~,\quad
16 \pi F_{1\perp}=  -\frac{8 g_{00} g_{rr}}{\sqrt{-g_{00}\prt{g_{rr}+x'^2 g_{11}}}}~.
\eea
The $F_{\perp}$ is independent of the anisotropic spatial direction metric elements which contributes exponential dilaton terms. It depends on the anisotropy through a combination of radial and metric elements only. The  $F_{\parallel}$  depends on $g_{33}$, and by observing that the $F_2$ term generated by the brane is the same in both expressions, we understand the stronger dependence on anisotropy of the anisotropic baryon versus the $SO(2)$ baryon.

\begin{figure}[ht]
\begin{minipage}[ht]{0.5\textwidth}
\begin{flushleft}
\centerline{\includegraphics[width=75mm]{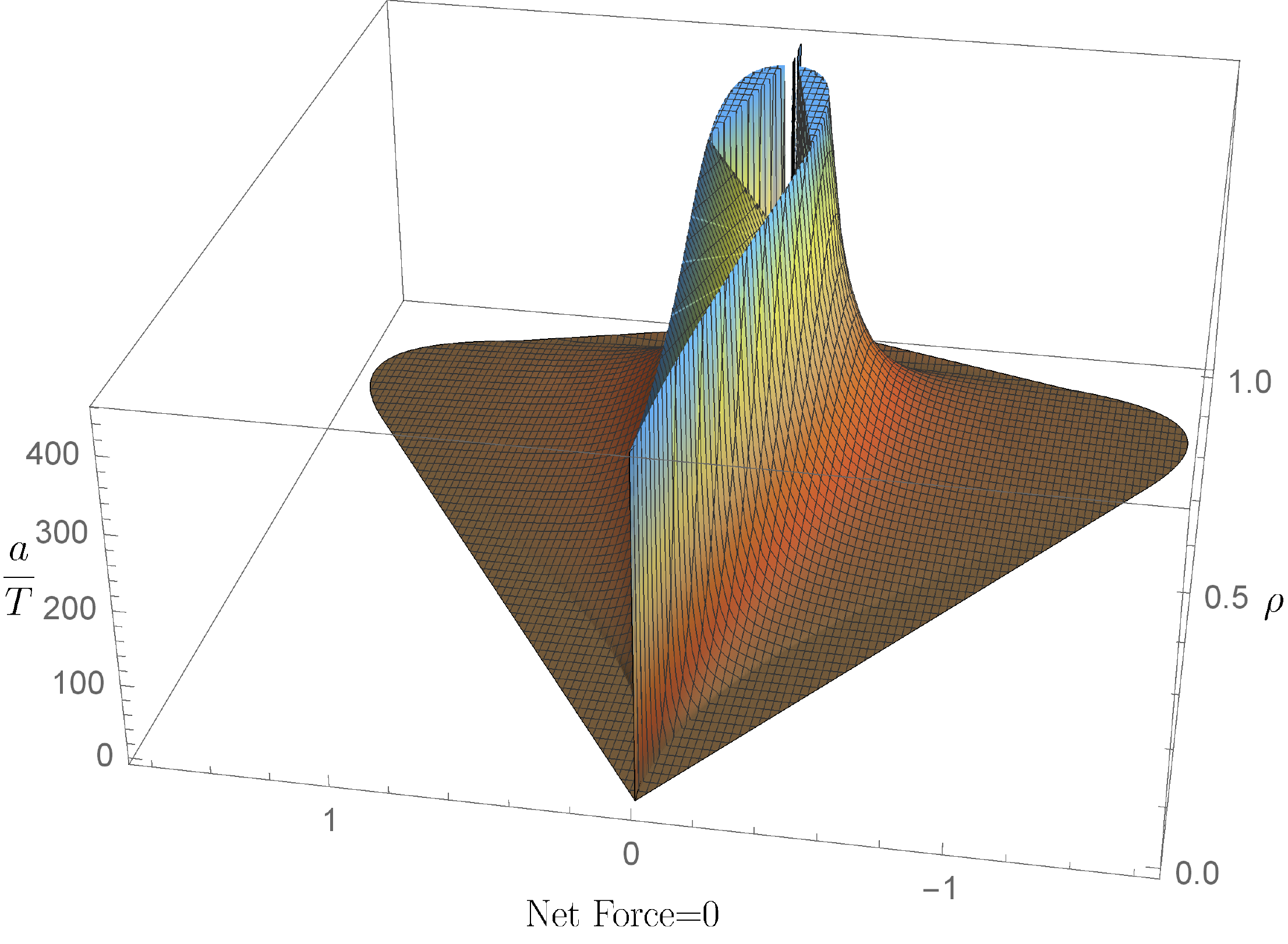}}
\caption{\small{The solutions of the net force  \eq{forcedp2} for the baryon on the anisotropic plane  with respect to the anisotropy for the complete range of holographic direction $\rho$. Notice that as the anisotropy is increased the surface tends quickly to saturate to a constant curve. The situation is similar to the rest of related quantities. It is the reason that at the high anisotropy regime, eventually the observables tend to be weakly dependent on the modifications of the anisotropy.}}
\label{fig:netf1}\vspace{0.7cm}
\end{flushleft}
\end{minipage}
\hspace{0.3cm}
\begin{minipage}[ht]{0.5\textwidth}
\begin{flushright}
\centerline{\includegraphics[width=90mm ]{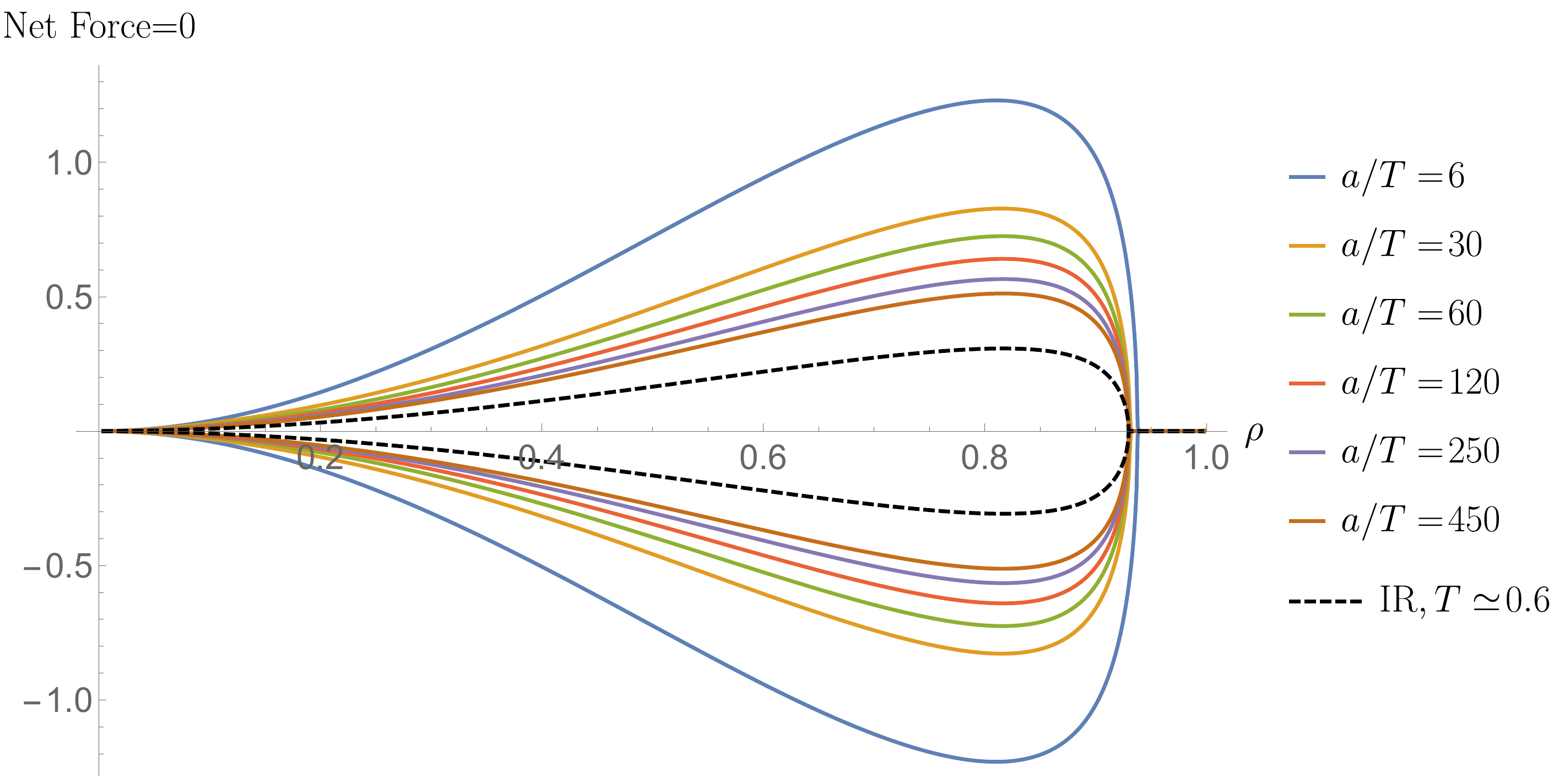}}
\caption{\small{The net force solution for a baryon vertex on the transverse plane for  fixed anisotropies, where the inner curves correspond to higher anisotropies, tending to converge. The dashed curve is for baryon in a holographic theory with background that matches IR region  \eq{azey1} of the RG flow of the Lifshitz-like background. The brane vertex is not in the deep IR of the RG  anisotropic flow background \eq{anisometric}, explaining the fact that  the baryon in the background \eq{azey1} has inner net force curve. The parameters are fixed such that the temperature of the Lifshitz-like space, matches the temperature of the RG flow at high anisotropies.} }
\label{fig:netf2}\vspace{-1.7cm}
\end{flushright}
\end{minipage}
\end{figure}

We find parametrically the solutions of the net force condition of the transverse baryon in the whole range of anisotropic and holographic space and present the result in the Figure \ref{fig:netf1}. It is easily demonstrated that the two dimensional surface tends to a particular curve with the increase of anisotropies. One may wonder why, and what determines the value of the curve?

The IR geometry is expected to give more dominant contributions to baryon in the high anisotropy regime. Nevertheless, the baryon vertex and the attached strings are not so deep in the bulk geometry to approximate the computation in the exact IR geometry. We check that by applying the formalism developed in  section \ref{section:generic} where we study the baryon on the Lifshitz-like theories.

In the IR the string frame metric \cite{Azeyanagi:2009pr} can be written in the following form
\be
ds^2_s=  r^2\prt{- f(r)dt^2+ dx_1^2
+dx_2^2}+ r^{\ff{24}{7}}dx_3^2+\ff{dr^2}{ a_1 r^\ff{12}{7}f(r)}
+b_1 r^\ff{2}{7} ds_{X_5}^2, \label{azey1}
\ee
where the dilaton, the blackening factor and the constants are
\be
e^{\phi}=b_1^2 r^\ff{4}{7}e^{\phi_0}, \qquad f(r)=1-\prt{\frac{r_h}{r}}^{\ff{22}{7}}~,\qquad a_1=\ff{49}{36}~ b_1^{-6}~ ,\qquad b_1=\prt{\ff{11}{12}}^\ff{1}{5}  \la{aze2}
\ee
and $\phi_0$ is a constant. The background is also and exact solution of type IIB supergravity.  We chose to omit all the details of the baryon analysis on this background, although some of them are interesting, and we present the relevant final results.

For a baryon on isotropic plane of the background \eq{azey1} the changes of anisotropy does not affect its properties. We compute the net force and then its solution for such a baryon to find
the results plotted in the Figure \ref{fig:netf2}, where we compare them with the solutions of the baryon in the anisotropic RG backgrounds \eq{anisometric}. The solutions of the RG background tend to inner curves as we increase the anisotropy, while the baryon in the background \eq{azey1} has inner net force curve explaining the fact that the baryon vertex in the RG background does not sit such deep in the bulk to approximate the dominant contributions from this region.  An interpretation of our findings is that some of the characteristics of the baryon in the RG theory, like the net force, the size and the energy, tend as the anisotropy increases towards the ones of the baryon in the theory \eq{azey1}. However this is difficult to strictly quantify since the strings are integrated over the holographic direction and probe the whole RG geometry, although for branes going deeper in the bulk, larger parts of the strings are subtracted by the regularization technique.

\section{Theories in Presence of Magnetic Fields and Baryons} \la{section:mag}

To demonstrate how the anisotropy is generated due to magnetic fields let us firstly consider the Einstein-Maxwell theory with a negative cosmological constant  as in \cite{DHoker:2009mmn}
\be\la{baction}
S=-\frac{1}{16 \pi G_5}\int dx^5\sqrt{-g}\prt{R+F^{MN} F_{MN} -\frac{12}{L^2}}+S_{GH}+ S_{CS}~,
\ee
where the Chern-Simons term ensure that the action correspond to the bosonic part of the $d=5$ minimal gauged supergravity and is useful to fix the normalization of the gauge field. Together with the Gibbons-Hawking term, these do not contribute to the background solution. Let us introduce a magnetic field
along the anisotropic $x_3$ direction with a field strength
\be
F= B dx_1 \wedge dx_2~.
\ee
By looking at the symmetries of the equations of motion of \eq{baction} it is obvious that it will introduce an anisotropy along the $x_3$ direction in the geometry as
\be\la{bmet}
ds^2=-f_0(r) dt^2 +\frac{dr^2}{f_0(r)}+f_1(r)\prt{dx_1^2+dx_2^2}+f_2(r)dx_3^2~.
\ee
Notice the space is anisotropic, and the particular solution of this model interpolates between $AdS_5$ and a geometry of BTZ black hole with $AdS_3\times R\times R$ isometries as
\be\la{bsol}
ds^2=-3 r^2\prt{1-\frac{r_h^2}{r^2}}dt^2+\frac{1}{3 r^2 \prt{1-\frac{r_h^2}{r^2}}} dr^2+\frac{1}{2 \sqrt{3}}|B|\prt{dx_1^2+dx_2^2} +3 r^2 dx_3^2~.
\ee
The background \eq{bmet} is anisotropic along the direction of the magnetic field $B$.   Here the internal part of the geometry is absent since we started with a 5-dimensional space. One may consider more involved geometries with the internal part of space and additional forms in string theory\footnote{A string theory background similar to magnetic brane solution is the non-commutative dual  theory, where our analysis can be also applied. The non-commutative field theories emerge as the world-volume theory of the D-branes with constant NS-NS B-field at certain limits on the parameters of the theory. In the string frame the backgrounds reads \cite{Hashimoto:1999ut,Maldacena:1999mh}
\bea\nn
&&ds^2= \a'R^2\prt{r^2 \prt{-f(r)dt^2+dx_1^2+h(r)\prt{dx_2^2+dx_3^2}}+\frac{dr^2}{r^2 f(r)}+d\Om_5^2}~,\\\nn
&&e^{2\phi}=g_s^2 h(r)~,\quad B_{23}= \a'R^2 a^2 r^4 h(r)~,\quad C_{01}=\frac{a^2 \a' R^2 r^4}{g_s}~,\quad F_{0123r}=\frac{4 \a'^2 R^4}{g_s}r^3 h(r)~,\\\nn
&&f(r)=1-\frac{r_h^4}{r^4}~,\quad h(r)=\frac{1}{1+a^4 r^4}~.
\eea
where $a$ is a constant parameter here related to the $\theta$ non-commutative parameter in field theory, $g_s$ the IR string coupling and $R^4=4\pi g_s N$ defined as usual.}. The exact expression of the additional forms, does not affect the qualitative details of our findings, due to the symmetries of the configuration.

\subsection{Baryon in Theories with Magnetic Fields}

In the previous section we have shown that magnetic fields induce anisotropic geometries of the form \eq{metric1} and \eq{anisometric}. Having presented already an extensive analytical and numerical analysis of the holographic baryons in the anisotropic theory with space-dependent $\theta$-term,
we like to focus here on the generic properties that the baryon expected to have in the presence of magnetic fields.

The DBI action for a generic magnetic background accommodating the backgrounds \eq{bmet} and \eq{bsol} reads
\be\la{dbiwz}
S_{D_p}=-T_p \int_{\cM_{p+1}}d^{p+1}\xi e^{-\phi}\sqrt{|G_{ab}+B_{ab}+2\pi\a' F_{ab}|}+ S_{WZ}~,
\ee
where F is the Born-Infeld field strength, and B is the NS-NS field, all of them computed on the induced geometry. The $S_{WZ}$ action contains the possible R-R forms of the background. The key point here is that the brane wraps the directions of the internal space and the time and since we integrate them, the above action will depend on the holographic direction $r$ for homogeneous field theories assuming that the brane solves the corresponding Euler-Lagrange equations. The position of the brane is static and is determined by the net force condition \eq{forcedp2}.  Therefore, the existence of additional R-R fields and the $B$-field in the DBI action modifies mainly the position of the brane on the holographic space.

Regarding the strings we have
\be
S_{NG}=\frac{1}{2\pi\a'}\int d\s d\t\prt{ \sqrt{-g}-\frac{1}{2} \e^{\a\b}B_{MN}\pp_\a X^\m \pp_\b X^N}~,
\ee
and it can be seen that by considering a constant magnetic field, the equations of motion are possible to be satisfied without introducing any  new qualitative features in the analysis.

Therefore, the model-independent features derived in the sections \ref{section:generic} and \ref{section:shape}, are features of the strongly coupled holographic theories in magnetic fields. The magnetic field  backreacts to the geometry and induces anisotropies, while it couples to the brane and string actions in a way already described, by affecting the baryon features only quantitative.  Namely the quark distribution of the baryon along the different directions, depends mainly on the anisotropy of the geometry triggered by the magnetic field and on the position of the Dp-brane. Moreover, baryons that live on any plane that is non-transverse to the magnetic field, will be dissociated in stages, as we have already explained. The relative magnitudes of
dissociation will be model dependent but the qualitative prediction is independent of the model.

Notice that the existence of the unified study scheme for heavy quark observables in anisotropic theories, which is insensitive to the source triggering the anisotropy is not so surprising. For example the general formalism for the heavy quark diffusion developed in \cite{Giataganas:2013hwa,Giataganas:2013zaa}
applies on certain limits of magnetic field environments \cite{Dudal:2018rki}.

\section{Effects of Strong Fields on Baryons with $k<N_c$ Quarks} \la{section:univ}

Let us consider a baryon with $k$-quarks, where $k$ is lower than the number of the colors $N_c$ in the
$SU(N_c)$ gauge theory. The holographic construction consists of $k$-fundamental strings initiating from boundary ending to the D5-brane. To stabilize the configuration and to ensure conservation of charges, we implement a number of $N_c-k$ strings initiating from the brane ending in the deep IR in the bulk of the theory.
We collect the boundary terms which give the no-force condition in the anisotropic background
\be\la{nof1}
k \frac{\pp L_{NG,1}}{\pp r'}- \prt{N_c-k}\frac{\pp L_{NG,2}}{\pp r'}= \frac{N_c}{4}\frac{\pp L_{DBI}}{\pp r}\Bigg|_{r=r_v}
\ee
where $L_{NG,1}:=\sqrt{\cD_1},~L_{DBI}:= \cD_2$ are the string and the brane parts respectively of the Lagrangian \eq{action1}, and $ L_{NG,2}:= \sqrt{-g_{00} g_{rr}}$ is the Lagrangian of the straight string initiating from the brane at $r_v$ to the deep bulk.

The equation \eq{nof1} reads
\be\la{nof2}
\frac{\sqrt{-g_{00} g_{rr}}}{\sqrt{\cD_1}}=\frac{N_c}{4 k}\frac{1}{\sqrt{-g_{00} g_{rr}}}\frac{\pp D_2}{\pp r} +\frac{N_c-k}{k}~
\ee
and constrains the number of strings according to
\be\label{kn1}
k=\frac{N_c\prt{\frac{B}{4}+1}}{\Gamma+1}~,
\ee
where we define $\Gamma$ to be equal to the LHS of the \eq{nof2} to notice that $\Gamma\le 1$,  and $B$ appears on the RHS of the same equation
\be
B:=\frac{1}{\sqrt{-g_{00} g_{rr}}}\frac{\pp D_2}{\pp r}=\frac{e^{-\phi}}{2 g_{00}\sqrt{g_{rr}}}g_{\th\th}^{\frac{3}{2}}\prt{5g_{00}g_{\th\th}'+g_{\th\th}\prt{g_{00}'-2g_{00}\phi'}}~.
\ee
Restricting the study to the class of theories with AdS asymptotics where the $\phi'$ and $g_{\th\th}'$ approach to zero faster than the space metric elements approach infinity, we find that the function $B$ is equal to the unit at the boundary. For the same flows close to the horizon in the deep IR the function scales as $1/\sqrt{u-u_h}$, therefore there $B$ becomes infinitely large. It would be very interesting to prove in general possible monotonicity of the function B by using holographic properties, like the c-theorem for anisotropic holography. By assuming theories with monotonic $B$ we obtain $B\ge 1$, with the minimum at the boundary and the maximum at the black hole horizon. By using the equation \eq{kn1} is not difficult to see that the baryon stability bound for such theories satisfies
\be\la{bound}
k\ge \frac{5}{8}N_c~,
\ee
and is the same with the AdS invariant theories.

For the theory with the space-dependent $\theta$-term  one can demonstrate this expectation  since the function $B$ is written as
\be
B=\frac{e^{\frac{\phi}{4}}}{2  \sqrt{g_{rr}}} \prt{\frac{1}{2} \phi +\log{g_{00}}}'.
\ee
By using the monotonicity of the metric elements and the dilaton, and the asymptotics of the background it can be proved that $B$ is an increasing function with minimum $\lim_{r\to \infty} B=1$ and maximum  $\lim_{r\to r_h} B =\infty$,  satisfying all the expectations mentioned above. Therefore the stability bound for the anisotropic baryon in this theory is given by \eq{bound}.

This is one of the few universal bounds of isotropic theories, that continue to be unmodified when anisotropic interactions appear. The main reason is that the relation \eq{bound} turns out to depend primarily on the metric element of the holographic direction and its asymptotics, where in the UV we have an $AdS$ space and in the IR at the horizon the metric element diverges. However for baryons in Lifshitz spacetimes, for example in the theory \eq{azey1}, the bound may be modified in a way that it depends on the critical exponent of the theory, and this may be an interesting further application of the methods developed here.

\section{Concluding Remarks} \label{section:final}

Our methodology and results have been presented in detail in the section \ref{section:method}. Here we briefly provide some concluding remarks. We have provided a generic formalism for the study of holographic baryons in strongly coupled anisotropic theories. Our formalism is applicable to several theories, including theories with space-dependent couplings and  theories under strong magnetic fields. We have also presented a full numerical analysis to quantify the qualitative expectations of baryons derived with the generic formalism we have developed. In anisotropic theories the baryon's quark distribution depends on the angle with respect to the direction that breaks the rotational symmetry, for example the magnetic field vector. For quarks lying on the anisotropic plane the distribution takes an elliptic shape to counterbalance the induced effect of gluon dynamics and retain a stable state. Such baryons dissociate in stages, depending on the angle of the quark with respect to the anisotropic direction, where certain pairs abandon the bound state first, followed by the closest pairs to them as the temperature increases.  For baryons with quarks less than the number of colors we have found that the universal stability condition, relating the numbers of quarks and the degree of the gauge group, remains unmodified in the theories under consideration. This is in contrast to several other independent universal relations, such as the shear viscosity over entropy density ratio, which tend to be violated in anisotropic theories. These are the main results of our work.

The feature of the baryon dissociation in stages we have observed is unique for static baryons although similar phenomena may be present in moving baryons in the plasma \cite{Athanasiou:2008pz}. Our observation may have potential qualitative applications on multi-quark bound states, for example, on the exotic baryons. Exotic states that are strongly coupled in the same way with the conventional baryons may dissociate in stages in presence of anisotropies with the pattern described,  which may not occur for example in molecule states. Our results could serve as a qualitative potential test to identify the nature of such states, assuming that the features observed at the large $N_c$ limit carry on for lower number of colors.

Our studies, due to the nature of gauge/gravity duality are limited to qualitative observations which nevertheless may turn out to be very insightful. Our baryons live in theories with large number of colors $N_c$, however this might not be an obstruction for the qualitative realization of our findings. Extensive simulations in 3 and 4-dim pure YM, support this claim for certain observables \cite{Teper:1998te,Teper:1998kw,Caselle:2000tn}.  In fact for the case of mesons the large $N_c$ computations have led to qualitative and semi-quantitative agreements for the masses of various states with the holographic results \cite{Bali:2013kia}. Regarding the baryons there are several techniques in lattice and effective field theories like the heavy baryon chiral perturbation theory, studying them in the framework of the  large $N_c$ limit and the $1/N_c$ expansion \cite{DeGrand:2012hd,Cordon:2014sda}. Therefore, it would be an interesting possibility to examine if any of the qualitative holographic large-$N_c$ baryon features observed in our studies  are realized in nature.

\section*{Acknowledgements}

The author acknowledges useful conversations with N. Irges, G. Leontaris, C. J. D. Lin, K. Zoubos. The author would also like to thank E. Arhontakis and 139 MM$\Pi$.  This work is supported by the grant 104-2112-M-007-001 -MY3 of the Ministry of Science and Technology of Taiwan.


\begin{appendices}

\section{Axion Deformed Gravity Dual Theory at Low Anisotropy Limit} \label{app:lowat}

At the low anisotropy, high temperature limit the metric \eq{anisometric} becomes
\bea\nn
&&\cF(r)=1 -\frac{r_h^4}{r^4}+a^2 \cF_2 (r)  +\mathcal{O}(a^4)~,\qquad \cB(r) = 1+a^2 \cB_2(r) +\mathcal{O}(a^4)~, \\
&&\cH(r)=e^{- \phi(r) }~,\qquad \phi(r) =  a^2\phi_2(r)+\mathcal{O}(a^4)~.
\la{lowae}
\eea
The unknown functions are determined by applying the usual horizon and boundary AdS conditions, to give
\bea
&&\cF_2(r)= \frac{ r_h^2}{24 }\left[ \frac{8}{r^2}( \frac{1}{r_h^2}-\frac{1}{r^2})- \frac{10}{r^4}\log 2 +(\frac{3}{r_h^4}+\frac{7}{r^4})\log\prt{1+\frac{r_h^2}{r^2}}\right]~,\\
&&B_2(r)=-\frac{1}{24 r_h^2}\left[\frac{10 }{\frac{r^2}{r_h^2}+1} +\log\prt{1+\frac{r_h^2}{r^2}}\right]~,\quad \phi_2(r)=-\frac{1}{4 r_h^2} \log\prt{1+\frac{r_h^2}{r^2}} ~,
\eea
while the horizon of the black hole is related to the ratio $a/T$ as
\be\label{uhteq}
T=\frac{r_h}{\pi} +a^2 \frac{5 \log2-2}{48 \pi r_h}+\mathcal{O}(a^4)~.
\ee
The backreaction of the axion at $\cO\prt{a^2}$ at the expression \eq{seca2} is given by
\bea\nn
X_{13}\prt{x'_b, \rho,\rt}&=& \int_{1}^{\infty} d\rt \frac{x'_b \rho \prt{1-\rho ^4}}{8 r_h  \prt{\r^4-\rt^4}^{1/2}}\prt{\frac{1}{\r^4\prt{\r^4-\rt^4}+r_h^4 x'_b{}^2(\rt^4-1)(\r^4-1)}}^{3/2}\cdot\\ \nn
&&
 \Bigg(\prt{\r^4-\rt^4}\prt{-2\r^4+r_h^4 x'_b{}^2(\r^4-1)}\log\prt{1+\r^2}+\bigg(2\r^4\prt{\r^4-\rt^4}+\\
&& r_h^4 x'_b{}^2\prt{2\rt^4-1-\r^4}\prt{\r^4-1}\bigg)\log\prt{1+\frac{\r^2}{\rt^2}}\Bigg)~. \label{x13lowat}
\eea

\end{appendices}

\bibliographystyle{JHEP}

\end{document}